\documentclass[11pt,a4paper,3p,final]{elsarticle}

\usepackage[utf8]{inputenc}
\usepackage{moreverb}
\usepackage{nomencl}
\usepackage{amsmath,amsfonts,bm}
\usepackage{verbatim}
\usepackage{algorithm,algpseudocode,setspace}
\usepackage[utf8]{inputenc}
\usepackage{soul}
\usepackage[dvipsnames]{xcolor}
\usepackage{import}

\usepackage[colorlinks]{hyperref}

\nomenclature{$a, A$}{scalars}
\newcommand{\T}[1]{\boldsymbol{#1}} \nomenclature{$\T{a},\T{A}$}{tensors}

\newcommand{\M}[1]{\underline{\T{#1}}}
\nomenclature{$\M{a},\M{A}$}{matrices}
\newcommand{\xspace}[1]{\mathbb{#1}}
\nomenclature{$\xspace{A}$}{finite-dimensional set}
\newcommand{\ispace}[1]{\mathcal{#1}}
\nomenclature{$\ispace{A}$}{infinite-dimensional space}

\newcommand\leftstar[1]{\hspace*{-.2em}~^\mathrm{tr}\hspace*{.01em}#1}

\newcommand{\trial}{\delta} \nomenclature{$\trial \bullet$}{trial quantity
$\bullet$}
\newcommand{\flc}{^*} \nomenclature{$\bullet\flc$}{fluctuating part of
$\bullet$}

\newcommand{\puc}{\Omega} \nomenclature{$\puc$}{periodic cell}
\newcommand{\de}[1]{\,{\mathrm d}#1}

\newcommand{\strain}{\varepsilon} \nomenclature{$\strain$}{strain}
 \nomenclature{$\strain$}{fluctuating
strain}

\nomenclature{$\strain$}{fluctuating strain}
\newcommand{\Tstrain}{\T{\strain}}
\newcommand{\Tstrainfl}{\Tstrain\flc}
\newcommand{\Ttstrainfl}{\trial\Tstrainfl}
\newcommand{\TE}{\T{E}}\nomenclature{$\TE$}{mean strain}
\newcommand{\ME}{\M{E}}

\newcommand{\stress}{\sigma} \nomenclature{$\stress$}{stress}
\newcommand{\Tstress}{\T{\stress}}

\newcommand{\sE}{\mathcal{E}}

\newcommand{\sH}{\mathcal{H}}
\newcommand{\sTN}{\mathcal{T}_{\T{N}}}

\newcommand{\xZ}{\mathbb{Z}}

\newcommand{\xZd}{\mathbb{Z}^{d}}
\newcommand{\xR}{\mathbb{R}}

\newcommand{\xC}{\mathbb{C}}

\newcommand{\xZNd}{\mathbb{Z}^2_{\T{N}}}
\newcommand{\xTN}{\mathbb{T}_{\TN}}
\newcommand{\xEN}{\mathbb{E}_{\TN}}

\newcommand{\imu}{\mathrm{i}}

\newcommand{\Mstrain}{\M{\varepsilon}}
\newcommand{\Mstrainfl}{\Mstrain^*}

\newcommand{\Mstress}{\M{\stress}}

\newcommand{\trn}{^\mathsf{T}}

\newcommand{\TN}{\T{N}}
\newcommand{\Tk}{\T{k}}
\newcommand{\Tm}{\T{m}}
\newcommand{\Tn}{\T{n}}
\newcommand{\Tx}{\T{x}}
\newcommand{\Ty}{\T{y}}
\newcommand{\TG}{\T{G}}
\newcommand{\aux}{^\mathrm{ref}}\nomenclature{$\bullet\aux$}{Quantity $\bullet$
related to the reference problem}
\newcommand{\TGamma}{\T{\Gamma}}\nomenclature{$\TGamma$}{Green function of the reference problem}
\newcommand{\MGamma}{\M{\Gamma}}
\newcommand{\Tf}{\T{\zeta}}
\newcommand{\Tt}{\T{\tau}}
\newcommand{\Tg}{\T{\theta}}
\newcommand{\Mf}{\M{\zeta}}
\newcommand{\Mt}{\M{\tau}}

\newcommand{\St}{\tau}

\newcommand{\gpoint}[1]{\T{x}_{\TN}^{#1}}
\newcommand{\FT}[1]{\widehat{#1}}
\newcommand{\bfun}[2]{\varphi^{#1}(#2)}
\newcommand{\meas}[1]{|#1|}
\newcommand{\bfunN}[2]{\varphi_{\TN}^{#1}(#2)}
\newcommand{\DFT}[2]{\omega_{\TN}^{#1#2}}

\newcommand{\xRNdd}{\xR^{3\meas{\TN}}}
\newcommand{\xCNdd}{\xC^{3\meas{\TN}}}

\newcommand{\gpw}{w}

\newcommand{\MF}{\M{F}}
\newcommand{\inv}{^{-1}}
\newcommand{\MG}{\M{G}}
\newcommand{\nite}{i}
\newcommand{\ite}[1]{_{(#1)}}
\newcommand{\MD}{\M{C}}
\newcommand{\norm}[1]{\| #1 \|}

\newcommand{\scontr}{\cdot}
\newcommand{\dcontr}{:}
\newcommand{\Tnablas}{\T{\nabla}_\mathrm{s}}
\nomenclature{$\Tnablas$}{Symmetrized gradient operator}

\newcommand{\TC}{\T{C}}\nomenclature{$\TC$}{Constitutive operator}
\newcommand{\TIs}{\T{I}_\mathrm{s}}\nomenclature{$\TC$}{Symmetic 4-order unit
tensor}
\newcommand{\Fz}[1]{\xi_{#1}(\Tk)} \nomenclature{$\Fz{i}$}{Reduced frequency
needed in Appendix A}

\newcommand{\etal}{\emph{et al.}}
\newcommand{\figref}[1]{Figure~\ref{#1}}
\newcommand{\appref}[1]{Appendix~\ref{#1}}
\newcommand{\tabref}[1]{Table~\ref{#1}}
\newcommand{\secref}[1]{Section~\ref{#1}}
\newcommand{\Eref}[1]{Eq.~\eqref{#1}}

\journal{arXiv}

\begin{document}

\begin{frontmatter}

\author[ctu]{J.~Zeman\corref{cor1}}
\ead{zemanj@cml.fsv.cvut.cz}
\ead[url]{http://mech.fsv.cvut.cz/~zemanj}
\cortext[cor1]{Corresponding author}
\author[m2i,tue]{T.W.J.de Geus}
\author[tub,ctu]{J. Vond\v{r}ejc}
\author[tue]{R.H.J. Peerlings}
\author[tue]{M.G.D. Geers}

\address[ctu]{Faculty of Civil Engineering, Czech Technical University in Prague, Th\'{a}kurova~7, 166~29 Praha~6, Czech Republic}
\address[m2i]{Materials Innovation Institute (M2i), P.O. Box 5008, 2600 GA Delft, The Netherlands}
\address[tub]{Institute of Scientific Computing, Technische Universität Braunschweig, D-38092 Braunschweig, Germany}
\address[tue]{Department of Mechanical Engineering, Eindhoven University of Technology, P.O. Box 513, 5600 MB Eindhoven, The Netherlands}

\title{A finite element perspective on non-linear FFT-based  micromechanical simulations}

\begin{abstract}
Fourier solvers have become efficient tools to establish structure-property relations in heterogeneous materials. Introduced as an alternative to the Finite Element (FE) method, they are based on fixed-point solutions of the Lippmann-Schwinger type integral equation. Their computational efficiency results from handling the kernel of this equation by the Fast Fourier Transform (FFT). However, the kernel is derived from an auxiliary homogeneous linear problem, which renders the extension of FFT-based schemes to non-linear problems conceptually difficult. This paper aims to establish a link between FE- and FFT-based methods, in order to develop a solver applicable to general history- and time-dependent material models. For this purpose, we follow the standard steps of the FE method, starting from the weak form, proceeding to the Galerkin discretization and the numerical quadrature, up to the solution of non-linear equilibrium equations by an iterative Newton-Krylov solver. No auxiliary linear problem is thus needed. By analyzing a two-phase laminate with non-linear elastic, elasto-plastic, and visco-plastic phases, and by elasto-plastic simulations of a dual-phase steel microstructure, we demonstrate that the solver exhibits robust convergence. These results are achieved by re-using the non-linear FE technology, with the potential of further extensions beyond small-strain inelasticity considered in this paper.
\end{abstract}

\begin{keyword}
periodic homogenization \sep FFT-based solvers \sep the Galerkin method \sep computational inelasticity \sep Newton-Krylov solvers
\end{keyword}
\end{frontmatter}

\section{Introduction}

The aim of computational micromechanics of materials is to establish a link between the mechanical response of two interacting scales in heterogeneous media, commonly referred to as the \emph{macro}- and \emph{micro}-scale. A pivotal role in this scale bridging is played by the \emph{local problem} -- a boundary value problem defined on a representative microscale sample that involves local constitutive laws, balance equations, and, most typically, periodic boundary conditions. The effective macroscopic response is then extracted from the solution of the local problem for a given macroscopic excitation, e.g.~\cite{michel_effective_1999,kanoute_multiscale_2009,geers_multi-scale_2010}.

For virtually all cases of practical relevance, the local problem must be solved approximately by discretizing the microstructure and the unknown microscopic fields. The prevailing technique employed for this purpose is the Finite Element method. However, the ever increasing desire to use finely discretized unit cells, even in 3D, calls for more efficient methods. In particular,  advances in experimental characterization of microstructures by high-resolution images triggers the need for efficient solvers that use these images directly as computational grids. A regular grid in combination with periodic boundary conditions naturally promotes solvers based on the Fast Fourier Transform~(FFT).

The first FFT-based numerical homogenization algorithm was proposed in the seminal work by Moulinec and Suquet as a suitable alternative to Finite Element methods~\cite{moulinec_fast_1994}.
In its original version, the method built on a fixed-point iterative solution of an \emph{integral equation} of the Lippmann-Schwinger type, whose kernel was derived from the Green function of a \emph{reference problem} -- an auxiliary local problem with a homogeneous constitutive law.
The efficiency and simplicity of the algorithm stems from the facts that (i)~the kernel is applied in the Fourier domain by optimized FFT routines (which are commonly available), (ii)~mesh generation is completely avoided through a use of the regular grid, and (iii)~the system/stiffness matrix does not have to be assembled.
Motivated by these attractive characteristics, several improvements of the basic scheme have been proposed to achieve a more robust performance~\cite{eyre_fast_1999,michel_computational_2000,michel_computational_2001,vinogradov_accelerated_2008,monchiet_polarization-based_2012}, eventually allowing the FFT-based algorithms to become a method of choice for multi-scale modeling of complex non-linear materials~\cite[and references therein]{montagnat_multiscale_2014,sliseris_numerical_2014,stein_fatigue_2014}.

Despite their over twenty-year history, the theoretical foundations of the FFT-based methods have been understood only recently, by distinguishing the \emph{discretization} from the solution of the resulting \emph{system of linear algebraic equations}.
In particular, Zeman\etal~\cite{zeman_accelerating_2010} found the integral formulation to be equivalent to a \emph{spectral collocation} method resulting in a fully populated system of linear equations with a sparse representation; the convergence
of approximate solutions for non-smooth coefficients has been proven by Vondřejc~\cite[pages 116--117]{Vondrejc2013PhD} and by Schneider~\cite{schneider_convergence_2014}. The original Moulinec-Suquet scheme is recovered when solving the system by the Richardson iteration~\cite{Mishra:2015:ACL}, but other low-memory iterative solvers, such as conjugate gradients, offer even better performance.
An alternative approach proposed by Brisard and Dormieux~\cite{brisard_fft-based_2010}, which was justified later~\cite{brisard_combining_2012,brisard_periodic_2014}, proceeds from the discretization of the \emph{Hashin-Shtrikman functional} with pixel/voxel-wise constant polarization fields, yielding again ``structurally sparse'' systems that can be efficiently treated by iterative solvers. Finally, Vondřejc and co-workers re-established the connection between FFT-based schemes and Finite Elements in the framework of conventional \emph{Galerkin methods} with a specific choice of basis functions and numerical quadrature \cite{Vondrejc2013PhD,vondrejc_fft-based_2014} or exact integration \cite{Vondrejc2015FFTimproved}. The main advantage of this approach is the fact that it does not rely on the notion of a reference problem. Such a feature is particularly attractive for non-linear problems, for which the concept of the Green functions cannot be used.

Building on recent theoretical results for linear problems~\cite{Vondrejc2013PhD,vondrejc_fft-based_2014}, this paper aims to explain and explore the close connection between the standard Finite Elements and FFT-based techniques in a non-linear setting. Our aim is to develop a robust, universal, and transparent Fourier formulation for non-linear and history-dependent constitutive laws in the small strain regime. In \secref{sec:variational} we cast FFT-based methods in the framework of standard non-linear Finite Element procedures and highlight many similarities, as well as a few differences. To simplify the explanation, the derivations are here based on non-linear elasticity. However, this treatment can be easily extended to arbitrary non-linear and history-dependent constitutive models through the well-known consistent tangent operators and time discretization schemes of computational inelasticity, e.g.~\cite{simo_computational_1998,neto_computational_2011}, as demonstrated in Sections~\ref{sec:model} and \ref{sec:Examples}. \secref{sec:connections} is devoted to a comparison of the proposed approach, the Finite Element Method, and non-linear FFT-based solvers available in the literature. The performance of the proposed method is demonstrated in \secref{sec:Examples} by analyzing a two-phase laminate with non-linear elastic, elasto-plastic, and visco-plastic phases; and finally a micrograph-based analysis of dual-phase steel. A summary is included in ~\secref{sec:conclusion}, along with possible extensions. Technical details are gathered in \ref{app:operators} and \ref{app:matrices}, in order to render the paper self-contained.

\section{Galerkin formulation}\label{sec:variational}

The purpose of this section is to derive, step by step, a non-linear FFT-based scheme in a setting parallel to Finite Element~(FE) formulations. The points of departure are the weak forms of the local problem (\secref{sec:weak}) and strain compatibility conditions (\secref{sec:compatibility}), under the small strain assumption. The latter represents the major difference between FE and FFT formulations. In \secref{sec:approx_space}, we introduce the approximation space, along with the properties of the basis functions that are required for the discretization of the weak form in \secref{sec:discretization}. The resulting system of non-linear nodal equilibrium equations is linearized in \secref{sec:linearization} leading to an incremental-iterative Newton-Krylov solution scheme outlined in \secref{sec:algorithm}.

The notation used is as follows. Scalar quantities are denoted by plain letters, e.g. $a$ or $A$. First-, second-, and fourth-tensors are in bold, e.g. $\T{a}$ or $\T{A}$~(where the rank will be clear from the context). The matrices arising from the discretization procedure are underlined, e.g. $\M{a}$ or $\M{A}$. To enhance readability, we limit ourselves to two dimensions under the plane strain assumption. Note, however, that the methodology is by no means restricted to 2D and the extension to higher dimensions is trivial.

\subsection{Local problem and its weak form}\label{sec:weak}

In what follows, we consider the microstructure of the material to be represented by a periodic cell $\puc = (-L_1/2, L_1/2) \times (-L_2/2, L_2/2)$ of area $\meas{\puc} = L_1 L_2$. The material response at a point $\Tx \in \puc$ is specified by the \emph{constitutive relation} $\Tstress(\Tx, \Tstrain(\Tx))$ assigning the stress response $\Tstress$ to a given strain $\Tstrain$ locally at $\Tx$. Furthermore, the total strain $\Tstrain$ is split into a homogeneous \emph{average strain} tensor $\TE$ and an $\puc$-periodic \emph{fluctuating strain} field $\Tstrainfl$, i.e.
\begin{align}
\Tstrain(\Tx) = \TE + \Tstrainfl(\Tx)
\text{ for }
\Tx \in \puc,
&&
\int_\puc
\Tstrainfl(\Tx)
\de\Tx
=
\T{0}.
\end{align}
The average strain $\TE$ represents a given macro-scale excitation, while the fluctuating micro-scale strain field $\Tstrainfl$ is the primary unknown.

The fluctuating strain field $\Tstrainfl$ is determined by the stress \emph{equilibrium} and strain \emph{compatibility} conditions, which under quasi-static assumptions and in small strains read as, e.g.~\cite[Section~3]{Milton:2002:TOC},
\begin{subequations}\label{eq:local_problem}
\begin{align}
- &
\T{\nabla} \scontr
\Tstress \big( \Tx, \TE + \Tstrainfl(\Tx) \big)
=
\T{0}
\text{ for }
\Tx \in \puc,
\label{eq:equilibrium}
\\
&
\Tstrainfl
\in \sE
=
\left\{
\Tnablas \T{u}\flc, \T{u}\flc \text{ is an }
\puc\text{-periodic displacement field}
\right\},
\label{eq:compatibility}
\end{align}
\end{subequations}
where $\T{\nabla} \cdot$ stands for the divergence operator and $\Tnablas$ stands for the symmetrized gradient operator. For the numerical treatment, the \emph{local problem}~\eqref{eq:equilibrium} is recast into the weak form, which amounts to finding $\Tstrainfl \in \sE$ such that
\begin{align}\label{eq:weak_form}
\int_\puc
\Ttstrainfl( \Tx ) \dcontr
\Tstress\big( \Tx, \TE + \Tstrainfl(\Tx) \big)
\de\Tx
=
0
\end{align}
holds for all $\Ttstrainfl \in \sE$~(where use has been made of the periodicity of the problem eliminate the boundary term).

\subsection{Compatibility}\label{sec:compatibility}

The main difference in how we proceed from the weak form~\eqref{eq:weak_form} with respect to the conventional FE method is in the way in which the compatibility constraint, Eq.~\eqref{eq:compatibility}, is imposed for both the solution $\Tstrainfl$ and the test fields $\Ttstrainfl$.
Commonly, these quantities are expressed with the help of $\puc$-periodic displacement fields $\T{u}\flc$ and $\delta \T{u}\flc$. As $\Tstrainfl = \Tnablas \T{u}\flc$ and $\Ttstrainfl = \Tnablas \delta \T{u}\flc$, their compatibility follows directly by definition~\eqref{eq:compatibility}, cf.~\secref{sec:fem}. Fourier-based methods, on the other hand, work directly with the strains and impose the compatibility of the solution and test fields by different means. For the test strains $\Ttstrainfl$, the compatibility is imposed via a projection operator $\TG$,
\begin{align}\label{eq:conv_def}
\Ttstrainfl( \Tx )
=
\big[ \TG \star \Tf \big] (\Tx)
=
\int_\puc \TG( \Tx - \Ty ) \dcontr \Tf(\Ty) \de \Ty
\text{ for }
\Tx \in \puc,
\end{align}
where $\star$ stands for the convolution. This operator maps an extended test function $\Tf$, taken from the space all of square-integrable symmetric tensor fields $\sH$, to its compatible part, i.e. $\TG \star \Tf \in \sE$ for all $\Tf \in \sH$. The compatibility of the solution, $\Tstrainfl \in \sE$, will be enforced by different means later in \secref{sec:linearization}.

The convolution format of
Eq.~\eqref{eq:conv_def} suggests that it can be conveniently treated using the Fourier transform, when the Fourier transform of the operator $\TG$ is known analytically. Indeed, direct application of the convolution theorem reveals that
\begin{equation}\label{eq:projection}
\left[ \TG \star \Tf \right] (\Tx)
=
\sum_{\Tk \in \xZ^2}
\FT{\TG}( \Tk )
\dcontr
\FT{\Tf}( \Tk ) \,
\bfun{\Tk}{\Tx}
\text{ for }
\Tx \in \puc,
\end{equation}
where $\Tk$ is the discrete frequency vector in the two-dimensional Fourier domain $\xZ^2$,
$\varphi^{\Tk}$ is the complex-valued Fourier basis function,
\begin{align}\label{eq:Fourier_basis}
\bfun{\Tk}{\Tx}
=
\exp \left(
2 \pi \imu
\left[
\frac{k_1 x_1}{L_1} + \frac{k_2 x_2}{L_2}
\right]
\right)
\text{ for }
\Tx \in \puc,
\end{align}
and $\FT{\Tf}(\Tk)$ stands the complex-valued Fourier transform of $\Tf(\Tx)$,
\begin{align}
\FT{\Tf}(\Tk)
=
\frac{1}{\meas{\puc}}
\int_\puc
\Tf( \Tx ) \,
\bfun{-\Tk}{\Tx}
\de \Tx
\text{ for }
\Tk \in \xZ^2.
\end{align}
\nomenclature{$\imu$}{Imaginary unit}%
\nomenclature{$\meas{\bullet}$}{Measure or cardinality of $\bullet$}%
The closed-form expression for the Fourier transform of the projection operator $\FT{\TG}$ is available in \appref{app:operators}, Eq.~\eqref{eq:FT_projection}, from which it follows that $\TG$ is a self-adjoint operator; see, e.g.,~\cite[Lemma~2]{vondrejc_fft-based_2014}. Notice that no approximation is made in~\eqref{eq:projection}, because all quantities are $\puc$-periodic and the sum is infinite.

Substituting~\eqref{eq:conv_def} into the weak formulation in Eq.~\eqref{eq:weak_form} and employing the self-adjointedness of $\TG$ provides an equivalent characterization of the unknown strain field $\Tstrainfl\in\sE$:
\begin{equation}\label{eq:weak_form2}
  \int_\puc
  \big[ \TG \star \Tf \big] (\Tx)
  \dcontr
  \Tstress\big( \Tx, \TE + \Tstrainfl(\Tx) \big)
  \de\Tx
  =
  \int_\puc
  \Tf(\Tx)
  \dcontr
  \big[ \TG \star \Tstress \big] \big( \Tx, \TE + \Tstrainfl(\Tx) \big)
  \de\Tx
  =
  0
\end{equation}
for all $\Tf \in \sH$. Because the extended test functions $\Tf$ are no longer constrained to be compatible, this form is better suited for the discretization than the original one in Eq.~\eqref{eq:weak_form}.\footnote{Note that the solution and the test functions now lie in different spaces, constrained $\sE$ and unconstrained $\sH$. Alternatively, one can work with the symmetric version and apply the projection in the last step, i.e. in \secref{sec:linearization}, similarly to \cite[Sections~5.2 and 5.3]{vondrejc_fft-based_2014}. Here, we decided to use the non-symmetric version because it renders the derivations more compact.}

\subsection{Basis functions}
\label{sec:approx_space}

\begin{figure}[h]
 \centering
 \def\svgwidth{.575\textwidth}
 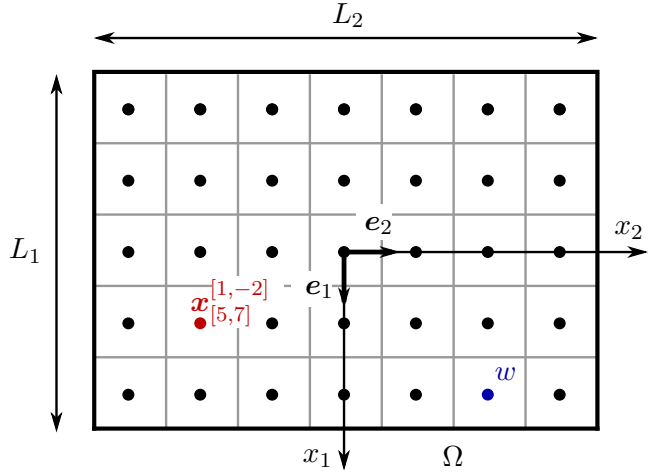
 \caption{An example of a $5 \times 7$ regular grid, $\xZ^2_{[5,7]}$,  discretizing the unit cell $\puc$ of dimensions $L_1 \times L_2$; the grid  nodes $\Tx^{\Tk}_{[5,7]}$ are indexed by $\Tk \in \xZ^2_{[5,7]}$. As an example $\bm{k} = [1,-2]$ is indicated in \textcolor[rgb]{0.78,0,0}{red}. Finally, $\gpw$ (indicated in \textcolor[rgb]{0.012,0,0.671}{blue}) stands for the nodal integration weight (equal to the pixel area).}
 \label{fig:finite_lattice}
\end{figure}

The basis functions rely on an underlying \emph{regular grid} with $\TN = [N_1, N_2]$ nodes along each coordinate, see \figref{fig:finite_lattice},
\nomenclature{$\TN$}{Discretization parameters}%
\begin{align}
\gpoint{\Tk}
=
\frac{k_1 L_1}{N_1} \T{e}_1
+
\frac{k_2 L_2}{N_2} \T{e}_2,
\end{align}
\nomenclature{$\gpoint{\bullet}$}{Node associated with index $\bullet$}%
on which the microstructure is sampled. The total number of the grid nodes is denoted as $\meas{\TN} = N_1 N_2$. As justified below, we shall consider only grids with an \emph{odd number} of nodes.

The individual nodes are indexed by a parameter $\Tk$ from a reduced index set
\begin{align}
\xZNd =
\left\{ \Tk \in \xZ^2,
- \frac{N_1}{2} < k_1 < \frac{N_1}{2},
- \frac{N_2}{2} < k_2 < \frac{N_2}{2}
\right\};
\end{align}
\nomenclature{$\xZNd$}{Truncated grid}%
it will become clear later that the indices $\Tk$ can be naturally identified with the discrete frequencies from~\eqref{eq:projection}. Finally, we assign the integration weight $\gpw = \meas{\puc} / \meas{\TN}$, equal to the pixel size, to each node.
\nomenclature{$\gpw$}{Weight of the Gauss points}

As follows from earlier developments~\cite{zeman_accelerating_2010,vondrejc_fft-based_2014}, it is convenient to use the \emph{fundamental trigonometric polynomials} defined on the grid $\xZNd$, e.g.~\cite[Chapter~8]{Saranen:2000:PIP},
\begin{align}\label{eq:bfunN}
\bfunN{\Tk}{\Tx}
=
\frac{1}{|\T{N}|}
\sum_{\Tm \in \xZNd}
\DFT{-\Tk}{\Tm}
\bfun{\Tm}{\Tx}
\text{ for }
\Tk \in \xZNd,
\end{align}
as the basis functions to approximate the weak form in Eq.~\eqref{eq:weak_form2}. Here, $\varphi^{\Tm}$ stands for the Fourier basis function (Eq.~\eqref{eq:Fourier_basis}) and $\DFT{\Tk}{\Tm}$ are the complex-valued coefficients of the \emph{Discrete Fourier Transform}~(DFT),
\begin{equation}\label{eq:DFT_def}
\DFT{\Tk}{\Tm}
=
\DFT{\Tm}{\Tk}
=
\bfun{\Tk}{\gpoint{\Tm}}
=
\exp\left(
2 \pi \imu
\left[ \frac{k_1 m_1}{N_1} + \frac{k_2 m_2}{N_2} \right]
\right)
\text{ for }
\Tk, \Tm \in \xZNd.
\end{equation}
\nomenclature{$\bfunN{\bullet}{\circ}$}{Fundamental trigonometric polynomial evaluated at the point $\circ$ associated with nodal point $\bullet$}%
\nomenclature{$\DFT{\bullet}{\circ}$}{Discrete Fourier Transform for indices
$\bullet$ and $\circ$}%

The solution $\Tstrainfl$ and the test functions $\Tf$ in Eq.~\eqref{eq:weak_form2} will be approximated as a linear combination of the basis functions $\varphi_{\TN}^{\Tk}$; the corresponding approximation space of the tensor-valued \emph{trigonometric polynomials} will be referred to as $\sTN$. These approximations are conforming, i.e. $\sTN \subset \sH$, as long as the number of nodes $\meas{\TN}$ is odd, e.g.~\cite[Section 4.3]{Vondrejc:2015:GUL}. This conformity is lost when $\meas{\TN}$ is even, resulting in a much more elaborate treatment, see~\cite[Section 4.4]{Vondrejc:2015:GUL}.
\begin{figure}[ht]
 \centering
 \includegraphics[width=1.\textwidth]{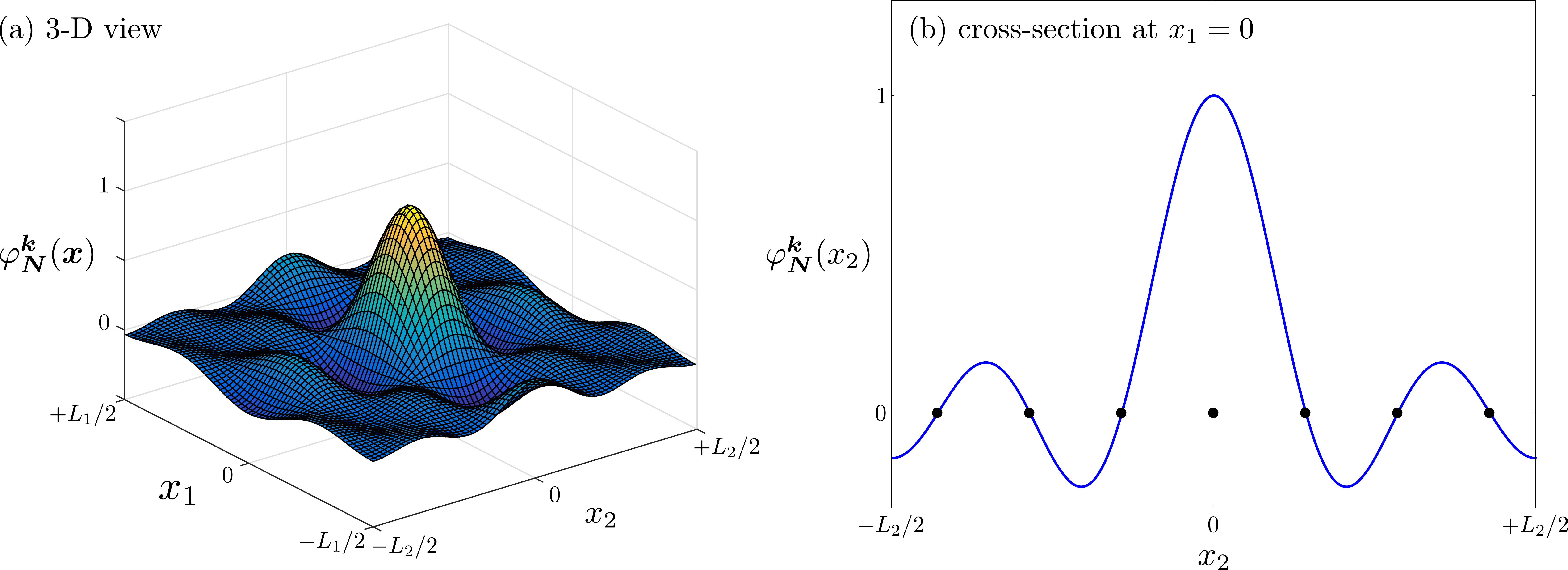}
 \caption{Example of a fundamental trigonometric polynomial, $\varphi_{\bm{N}}^{\bm{k}}$ with $\bm{N} = [5,7]$ and $\Tk = [0,0]$; as (a) a 3-D view and (b) a cross-section at $x_1 = 0$,  where the nodes are indicated with markers.}
 \label{fig:fund_trig_poly}
\end{figure}

The computational convenience of trigonometric polynomials follows from the fact that they can be efficiently manipulated using the Fast Fourier Transform~(FFT)~\cite{cooley_algorithm_1965}, because of (i)~the involvement of the DFT coefficients $\DFT{\Tk}{\Tm}$ in Eq.~\eqref{eq:bfunN} and (ii)~the ability to work with quantities defined in the Fourier space, because they incorporate the Fourier basis functions $\varphi^{\Tm}$. In the remainder of this section, we collect the most important steps needed to discretize the weak form in Eq.~\eqref{eq:weak_form2}; additional details are available e.g. in~\cite{Saranen:2000:PIP,Boyd:2001:CFS,vondrejc_fft-based_2014}. The reader familiar with trigonometric polynomials may proceed directly to the discretization procedure in \secref{sec:discretization}.

As can be seen from \figref{fig:fund_trig_poly}, in the real space the fundamental trigonometric polynomials are not locally supported, unlike the conventional Finite Element shape functions, however they are still \emph{interpolatory} and form the \emph{partition-of-unity}, because they satisfy
\begin{align}\label{eq:Kronecker_delta}
\varphi_{\T{N}}^{\T{k}}(\T{x}^{\T{m}}_{\T{N}})
=
\delta^{\T{k}\T{m}}
\text{ for }
\Tk, \Tm \in \xZNd,
&&
\sum_{\Tk \in \xZNd}
\varphi_{\T{N}}^{\T{k}}( \Tx )
= 1
\text{ for } \Tx \in \puc,
\end{align}
where $\delta^{\T{k}\T{m}}$ is the Kronecker delta. In the Fourier domain, they are
locally supported on $\xZNd$,
\begin{align}\label{eq:zero_frequencies}
\FT{\varphi}_{\TN}^{\Tk}( \Tm )
=
0
\text{ for }
\Tk \in \xZNd, \Tm \in \xZ^2 \backslash \xZNd,
\end{align}
because their definition (Eq.~\eqref{eq:bfunN}) contains only the Fourier basis functions $\varphi^{\Tm}$ associated with the frequencies from the grid $\xZNd$.

As a consequence, every trigonometric polynomial $\Tt \in \sTN$ admits two equivalent representations on the same grid $\xZNd$ that involve its nodal values $\Tt(\gpoint{\Tk})$, and the Fourier coefficients $\FT{\Tt}(\Tk)$. Their mutual relation is established by the forward and inverse DFTs,
\begin{align}\label{eq:DFT_forward_inverse}
\FT{\Tt}(\Tk)
=
\frac{1}{\meas{\TN}}
\sum_{\Tm \in \xZNd}
\DFT{-\Tk}{\Tm} \,
\Tt(\gpoint{\Tm}),
&&
\Tt(\gpoint{\Tk})
=
\sum_{\Tm \in \xZNd}
\DFT{\Tk}{\Tm} \,
\FT{\Tt}(\Tm)
\text{ for }
\Tk \in \xZNd.
\end{align}

\paragraph{Numerical integration} The scalar product of two trigonometric polynomials $\Tt \in \sTN$ and $\Tg \in \sTN$ can be evaluated \emph{exactly} by the trapezoidal rule,
\begin{align}\label{eq:trapz}
\int_{\puc}
\Tt(\Tx) \dcontr \Tg(\Tx)
\de\Tx
=
\gpw
\sum_{\Tk \in \xZNd}
\Tt(\gpoint{\Tk}) \dcontr \Tg(\gpoint{\Tk}),
\end{align}
which assigns the same integration weight, equal to the pixel area $\gpw$, to each grid node.

\paragraph{Convolution} of a trigonometric polynomial $\Tt \in \sTN$ with the projection operator $\TG$ from~\eqref{eq:conv_def} can be evaluated efficiently at the grid nodes $\gpoint{\Tk}$ by DFT. Indeed, a direct calculation reveals that
\begin{align}
\big[ \TG \star \Tt \big] (\gpoint{\Tk})
& \overset{\eqref{eq:projection}}{=}
\sum_{\Tm \in \xZd}
\FT{\TG}( \Tm )
\dcontr
\left[\,
\FT{\Tt}( \Tm ) \,
\bfun{\Tm}{\gpoint{\Tk}}
\right]
\overset{(\ref{eq:DFT_def}, \ref{eq:zero_frequencies})}{=}
\sum_{\Tm \in \xZNd}
\FT{\TG}( \Tm )
\dcontr
\left[\,
\FT{\Tt}( \Tm ) \,
\DFT{\Tk}{\Tm}
\right]
\nonumber \\
& \overset{\eqref{eq:DFT_forward_inverse}_1}{=}
\sum_{\Tm \in \xZNd}
\DFT{\Tk}{\Tm} \,
\FT{\TG}( \Tm )
\dcontr
\Bigg[
\frac{1}{\meas{\TN}}
\sum_{\Tn \in \xZNd}
\DFT{-\Tm}{\Tn} \,
\Tt(\gpoint{\Tn})
\Bigg]
\nonumber \\
& =
\sum_{\Tm \in \xZNd}
\sum_{\Tn \in \xZNd}
\Bigg[
\frac{1}{\meas{\TN}}
\DFT{\Tk}{\Tm} \,
\FT{\TG}( \Tm ) \,
\DFT{-\Tm}{\Tn}
\Bigg]
\dcontr
\Tt(\gpoint{\Tn})
\text{ for }
\Tk \in \xZNd.
\label{eq:projection_xk}
\end{align}

\paragraph{Matrix representation} All operations above only involve the discrete values at the grid $\xZNd$ in the real and in the Fourier spaces. It is therefore useful to employ a matrix representation, in which the column matrices
\begin{align}\label{eq:matrix_representation_def}
\Mt = \Big[\, \Mt^{\Tk} \,\Big]_{\Tk \in \xZNd}
\text{ with }
\Mt^{\Tk} = \Tt ( \gpoint{\Tk} ),
&&
\FT{\Mt} = \Big[\, \FT{\Mt}^{\Tk} \,\Big]_{\Tk \in \xZNd}
\text{ with }
\FT{\Mt}^{\Tk} = \FT{\Tt}( \Tk ),
\end{align}
collect the values of the trigonometrical polynomial $\Tt$ and its Fourier transform $\FT{\Tt}$ on the grid $\xZNd$. The one-to-one map between $\Mt$ and $\FT{\Mt}$,
\begin{align}\label{eq:DFT_matrices}
\FT{\Mt} = \MF \, \Mt, &&
\Mt
=
\MF\inv \, \FT{\Mt}
\end{align}
is established with the help of complex-valued matrices $\MF$ and $\MF\inv$ implementing the forward and inverse tensor-valued DFT according to~\eqref{eq:DFT_forward_inverse}.

In this matrix notation, the projection~\eqref{eq:projection_xk} attains the form
\begin{align}\label{eq:discrete_projection}
\Bigl[ \big[ \TG \star \Tt \big] (\gpoint{\Tk}) \Bigr]_{\Tk \in \xZNd}
&=
\MF\inv \, \FT{\MG} \, \MF \, \Mt
=
\MG \, \Mt,
\end{align}
where the real-valued matrix $\MG$ is symmetric, because the DFT matrices satisfy $\MF\inv = \meas{\TN} \MF^\mathrm{H}$ with $\mathrm{H}$ denoting the complex (Hermitian) transpose. The crux of the computational efficiency of Fourier-based methods is that the multiplication with $\MG$ is fast, because the action of $\MF$ and $\MF\inv$ can be efficiently implemented with FFT and $\FT{\MG}$ is block-diagonal in Fourier space. These properties are clarified in~\appref{app:matrices}, where the matrix notation is elaborated in full detail.
\nomenclature{$\MG$}{Discrete projection operator}

\subsection{Discretization}
\label{sec:discretization}

Now we are in the position to discretize the weak form of Eq.~\eqref{eq:weak_form2} with trigonometric polynomials. Following the standard Galerkin procedure, we approximate the unknown field $\Tstrainfl$ and the test field $\Tf$ in the same way:
\begin{subequations}\label{eq:approximation_1}
\begin{align}
\Tstrainfl( \Tx )
& \approx
\sum_{\Tm \in \xZNd}
\bfunN{\Tm}{\Tx} \,
\Tstrainfl( \gpoint{\Tm} )
\stackrel{\eqref{eq:matrix_representation_def}}{=}
\sum_{\Tm \in \xZNd}
\bfunN{\Tm}{\Tx} \,
{\Mstrainfl}^{\Tm},
\label{eq:approximation_1a}
\\
\Tf( \Tx )
& \approx
\sum_{\Tm \in \xZNd}
\bfunN{\Tm}{\Tx} \,
\Tf( \gpoint{\Tm} )
\stackrel{\eqref{eq:matrix_representation_def}}{=}
\sum_{\Tm \in \xZNd}
\bfunN{\Tm}{\Tx} \,
\Mf^{\Tm}.
\end{align}
\end{subequations}
The column matrices of nodal strains $\Mstrainfl$ and of nodal values of test fields $\Mf$ are respectively located in the corresponding finite-dimensional spaces $\xEN \subset \xTN$. The (constrained) space $\xEN$ thus collects the nodal values of \emph{compatible} trigonometric polynomials from $\sTN \cap \sE$, whereas (unconstrained) $\xTN$ collects nodal values of \emph{all} trigonometric polynomials from $\sTN$, see \Eref{eq:discrete_spaces} from Appendix~\ref{app:matrices} for details.

Introducing these expansions into~\eqref{eq:weak_form2} provides the condition for the nodal values of strain fields~$\Mstrainfl$,
\begin{align}\label{eq:weak_form_discretized}
\int_\puc
\Bigl(
\sum_{\Tm \in \xZNd}
\bfunN{\Tm}{\Tx} \,
\Mf^{\Tm}
\Bigr)
:
\Big[ \TG \star \Tstress \Big]
\Bigl(
\Tx,
\TE
+
\sum_{\Tm \in \xZNd}
\bfunN{\Tm}{\Tx} \,
{\Mstrainfl}^{\Tm}
\Bigr)
\de\Tx
=
0,
\end{align}
to be satisfied for arbitrary $\Mf$ from $\xTN$.

Application of the trapezoidal quadrature rule~\eqref{eq:trapz} provides
\begin{align}\label{eq:discretization_1}
\gpw
\sum_{\Tk \in \xZNd}
\Bigl(
\sum_{\Tm \in \xZNd}
\bfunN{\Tm}{\gpoint{\Tk}} \,
\Mf^{\Tm}
\Bigr)
:
\Big[ \TG \star \Tstress \Big]
\Bigl(
\gpoint{\Tk},
\TE
+
\sum_{\Tm \in \xZNd}
\bfunN{\Tm}{\gpoint{\Tk}} \,
{\Mstrainfl}^{\Tm}
\Bigr)
\approx
0;
\end{align}
note that this step introduces an approximation error because the constitutive relation $\Tstress$ does not necessarily map trigonometric polynomials to trigonometric polynomials. By exploring the Kronecker delta property of the basis functions~\eqref{eq:Kronecker_delta}$_1$, the previous relation further simplifies to
\begin{align}\label{eq:discretization_2}
\sum_{\Tk \in \xZNd}
\Mf^{\Tk}
:
\big[ \TG \star \Tstress \big]
\bigl(
\gpoint{\Tk},
\ME
+
{\Mstrainfl}^{\Tk}
\bigr)
=
0.
\end{align}

The discretization procedure is completed by employing the matrix representation of the projection operator~\eqref{eq:discrete_projection}, which transforms~\eqref{eq:discretization_2} into
\begin{align}\label{eq:discretization_3}
\Mf \trn \,
\MG \,
\Mstress
\left( \ME + \Mstrainfl \right)
=
0
\text{ for all } \Mf \in \xTN.
\end{align}
Here, $\Mstress$ denotes the constitutive law evaluated \emph{locally} at the grid nodes,
\begin{align}\label{eq:matrix_constitutive_law}
\Mstress( \ME + \Mstrainfl )
=
\left[
\Tstress
\bigl(
\gpoint{\Tk},
\ME
+
{\Mstrainfl}^{\Tk}
\bigr)
\right]_{\Tk \in \xZNd}.
\end{align}
Because the test matrices $\Mf$ are arbitrary, we finally distill from~\eqref{eq:discretization_3} that the nodal strain values $\Mstrainfl \in \xEN$ follow from the system of non-linear \emph{nodal equilibrium} conditions,
\begin{align}\label{eq:nodal_equilibrium_equations}
\MG \,
\Mstress(\ME + \Mstrainfl)
=
\M{0},
\end{align}
where the non-linearity originates solely from the constitutive relation,
because the projection matrix $\MG$ is independent of $\Mstrainfl$. Therefore, apart from enforcing the strain compatibility, the symmetric matrix $\MG$ also enforces the nodal equilibrium conditions, cf.~\cite[Lemma~2]{vondrejc_fft-based_2014}. Also notice that, in analogy to \secref{sec:compatibility}, the constraint $\Mstrainfl \in \xEN$ still needs to be accounted for.

\subsection{Linearization}\label{sec:linearization}

The conventional Newton scheme is used to find the solution to the system~\eqref{eq:nodal_equilibrium_equations} iteratively. For this purpose, we express the nodal unknowns in the $(\nite+1)$-th iteration as
\begin{align}
\Mstrainfl\ite{\nite+1}
=
\Mstrainfl\ite{\nite}
+
\delta\Mstrainfl\ite{\nite+1},
\end{align}
\nomenclature{$\nite$}{Iterations of the Newton method}
\nomenclature{$\ite{\bullet}$}{Value at the $\bullet$ iterate}
and linearize~\eqref{eq:nodal_equilibrium_equations} around $\Mstrainfl\ite{\nite}$, with $\Mstrainfl\ite{0} \in \xEN$. As a result, we obtain the linear system for the nodal strain increment $\delta\Mstrainfl\ite{\nite + 1} \in \xEN$:
\begin{align}\label{eq:linear_system}
\MG \,
\MD\ite{\nite}
\delta\Mstrainfl\ite{\nite + 1}
=
-
\MG \, \Mstress(\ME + \Mstrainfl\ite{i}),
\end{align}
\nomenclature{$\MD$}{Consistent constitutive tangent}%
where the tangent matrix
\begin{align}
\MD\ite{\nite}
=
\frac{\partial \Mstress}{\partial \Mstrainfl}
\left( \ME + \Mstrainfl\ite{\nite} \right)
\end{align}
is block-diagonal, by the locality of the stress-strain map~\eqref{eq:matrix_constitutive_law}, and its $\Tk$-th block is given by
\begin{equation}\label{eq:local_tangent}
\MD^{\Tk}\ite{\nite}
=
\frac{\partial \Tstress}{\partial \Tstrain}
\left( \gpoint{\Tk}, \TE + \Mstrain\ite{\nite}^{*\Tk} \right)
\text{ for }
\Tk \in \xZNd,
\end{equation}
see again \appref{app:matrices} for details. This matrix thus collects \emph{local} constitutive tangents evaluated \emph{independently} at the nodes.

Three considerations must be taken into account when solving the linearized system~\eqref{eq:linear_system}: (i)~the system matrix is \emph{dense}, \emph{singular}, and very costly to assemble for large    grids, (ii)~the multiplication with the system matrix is cheap and does not require the matrix assembly, because it involves the multiplication with \emph{structurally sparse} matrices~(recall that the multiplication with $\MG$ can be performed efficiently by FFT, Eq.~\eqref{eq:discrete_projection}, and $\MD\ite{\nite + 1}$ is block-diagonal), and (iii)~the solver must enforce the compatibility constraint $\Mstrainfl\ite{\nite + 1} \in \xEN$. All these aspects invite the application of (projected) iterative solvers involving only matrix-vector products, such as specific-purpose solvers~\cite{Moulinec2014comparison}, or selected general-purpose iterative algorithms for symmetric positive systems~\cite{Mishra:2015:ACL}, because the projection matrix $\MG$ enforces the compatibility and equilibrium conditions simultaneously. Specifically, we will use the conventional Conjugate Gradient algorithm~\cite{Hestenes:1952:MCG}, which  enforces the compatibility constraint at every iteration and outperforms alternative solvers in terms of convergence rate, as demonstrated recently in~\cite{Mishra:2015:ACL}.

\subsection{Algorithm}\label{sec:algorithm}

To summarize, the incremental-iterative Newton--Conjugate Gradient solver is outlined as a pseudo-algorithm in Algorithm~\ref{fig:algorithm}. We
emphasize for later reference that the algorithm implements two termination criteria for the Newton~(line 7) and the Conjugate Gradient~(line 9) solvers that involve the two tolerances $\eta^\mathrm{NW}$ and $\eta^\mathrm{CG}$, respectively.
Finally, note that the same procedure applies to history- and rate-dependent material laws, once the time-incremental stress-strain laws and consistent constitutive tangents are adopted, replacing $\underline{\bm{\sigma}}_{(i)}$ and $\MD\ite{\nite}$ in Eq.~\eqref{eq:linear_system} and lines 8 and 9 the algorithm. See, e.g.,~\cite{simo_computational_1998,neto_computational_2011} for a general treatment of such constitutive laws and \secref{sec:model} for specific examples.

\begin{algorithm}[ht]
\caption{Pseudo-algorithm of the variational FFT method}
\label{fig:algorithm}
\setstretch{2.2}
\begin{algorithmic}[1]
  \vspace*{1em}
  \State
  \(
    t = t_0
  \)
  \Comment\scriptsize
  Initial conditions
  \State
  \(
    \underline{\bm{\varepsilon}}^\star_{(0)} =
    \underline{\bm{0}}
  \)
  \Comment\scriptsize
  No fluctuations
  \State
  \(
    \ldots
  \)
  \Comment\scriptsize
  \textit{Initialize other history variables (material dependent)}
  \While{%
  \(
    t \leq T
  \)}
  \Comment\scriptsize
  \textbf{\underline{(i) Increment loop}}
  \State
  \(
    i = 0
  \)
  \Comment\scriptsize
  Reset iteration counter
  \State
  \(
    \delta \underline{\bm{\varepsilon}}^\star_{(i)} =
    \underline{\bm{\infty}}
  \)
  \Comment\scriptsize
  Initialize, indicating no convergence yet
  \While{%
  \(
    \big|\big|\, \delta \underline{\bm{\varepsilon}}^\star_{(i)} \,\big|\big| \, / \,
    \big|\big|\, \underline{\bm{E}}_{(t)} \,\big|\big| > \eta^\text{NW}
  \)}
  \Comment\scriptsize
  \textbf{\underline{(ii) Newton loop}}
  \State
  \(
    \underline{\bm{\sigma}}_{(i)} \,
    {\underline{\bm{\sigma}}_{(i)}} \,
    {\underline{\bm{C}}_{(i)}}
    =
    \;\;\underline{\bm{\sigma}}
    {\;\;\underline{\bm{\sigma}}} \,
    {\displaystyle\frac{\partial \underline{\bm{\sigma}}}{\partial \underline{\bm{\varepsilon}}}}
    \big(\, \underline{\bm{E}}_{(t)} +  \underline{\bm{\varepsilon}}^\star_{(i)} \,\big)
  \)
  \Comment
  Constitutive response \textit{(material dependent)}
  \State
  \(
    \underline{\bm{C}}_{(i)} =
    \displaystyle\frac{\partial \underline{\bm{\sigma}}}{\partial \underline{\bm{\varepsilon}}}
    \big(\, \underline{\bm{E}}_{(t)} + \underline{\bm{\varepsilon}}^\star_{(i)} \,\big)
  \)
  \Comment\scriptsize
  Consistent tangent \textit{(material dependent)}
  \While{%
  \(
    \big|\big|\,
      \underline{\bm{G}} \,
      \underline{\bm{C}}_{(i)} \,
      \delta \underline{\bm{\varepsilon}}^\star_{(i+1)} +
      \underline{\bm{G}} \,
      \underline{\bm{\sigma}}_{(i)} \,
    \big|\big|\,
    / \,
    \big|\big|\,
      \underline{\bm{G}} \,
      \underline{\bm{\sigma}}_{(i)} \,
    \big|\big|\,
    > \eta^\text{CG}
  \)}
  \Comment\scriptsize
  \textbf{\underline{(iii) Iterative linear solver\vphantom{p}}}
  \State
  \ldots
  \Comment\scriptsize
  Standard Conjugate Gradients, for: %
  \(
    \underline{\bm{G}} \,
    \underline{\bm{C}}_{(i)} \,
    \delta \underline{\bm{\varepsilon}}^\star_{(i+1)} =
    - \underline{\bm{G}} \, \underline{\bm{\sigma}}_{(i)}
  \)
  \EndWhile
  \State
  \(
    \underline{\bm{\varepsilon}}^\star_{(i+1)} =
    \underline{\bm{\varepsilon}}^\star_{(i)} +
    \delta \underline{\bm{\varepsilon}}^\star_{(i+1)}
  \)
  \Comment\scriptsize
  Iterative update
  \State
  \(
    i = i+1
  \)
  \Comment\scriptsize
  Proceed to next Newton iteration
  \EndWhile
  \State
  \(
    \underline{\bm{\varepsilon}}^\star_{(t+\Delta t)} =
    \underline{\bm{\varepsilon}}^\star_{(i)}
  \)
  \Comment\scriptsize
  ``Initial guess'' for the next increment
  \State
  \(
    \ldots
  \)
  \Comment\scriptsize
  \textit{Update other history variables (material dependent)}
  \State
  \(
    t = t + \Delta t
  \)
  \Comment\scriptsize
  Proceed to next increment
  \EndWhile
  \vspace*{1em}
\end{algorithmic}
\setstretch{1.0}
\end{algorithm}

\section{Connections to other methods}\label{sec:connections}

\subsection{Finite elements}\label{sec:fem}

We have demonstrated in \secref{sec:variational} that the presented formulation of FFT-based methods shares many similarities with Finite Element~(FE) methods, such as the Galerkin discretization procedure, numerical quadrature, or linearization of nodal equilibrium conditions. However, it deviates in (i)~enforcing compatibility of the solution and of the test fields, and in (ii)~the choice of basis functions. In the current section, we investigate the implications of these two differences in more detail.

Specifically, the point of departure of the FE discretization is the weak formulation of the local problem~\eqref{eq:weak_form}, expressed in terms of displacement fluctuations $\T{u}\flc$, e.g.~\cite{michel_effective_1999}:
\begin{align}
\int_\puc
\Tnablas
\delta \T{u}\flc( \Tx )
\dcontr
\Tstress\big( \Tx, \TE + \Tnablas
\T{u}\flc( \Tx ) \big)
\de\Tx
=
0,
\end{align}
where both the solution $\T{u}\flc$ and the test function $\delta \T{u}\flc$ are $\puc$-periodic displacement fields, whose mean is set to zero to eliminate the rigid body modes.

Applying the standard FE technology, e.g.~\cite{Bathe:1996:FEP}, we find that the nodal values of the displacement fluctuations $\M{u}\flc$ follow from the non-linear system of nodal equilibrium equations
\begin{align}\label{eq:nodal_equil_FE}
\sum_{g = 1}^{n}
\gpw^g
\M{B}\trn(\Tx^g)
\Tstress
\big( \Tx^g, \TE + \M{B}( \Tx^g ) \M{u}\flc
\big)
=
\M{0},
\end{align}
where $\Tx^g$ refers to the positions of $n$ Gauss integration points, $w^g$ are their weights, and $\M{B}$ stands for the symmetrized gradient of the Lagrange basis functions. The non-linear system~\eqref{eq:nodal_equil_FE} is typically solved iteratively by the Newton method, which, following the steps and the notations of \secref{sec:linearization}, yields the following linear system for the nodal iterative displacement update $\delta\M{u}\flc\ite{\nite+1}$:
\begin{align}\label{eq:linear_system_FE}
\left[
\sum_{g = 1}^{n}
\gpw^g
\M{B}\trn(\Tx^g)
\frac{\partial \Tstress}{\partial \Tstrain}
\bigl( \Tx^g, \TE + \M{B}( \Tx^g ) \M{u}\ite{\nite}\flc
\bigr)
\right]
\delta\M{u}\flc\ite{\nite+1}
=
-
\sum_{g = 1}^{n}
\gpw^g
\M{B}\trn(\Tx^g)
\Tstress
\bigl( \Tx^g, \TE + \M{B}( \Tx^g ) \M{u}\ite{\nite}\flc
\bigr)
\end{align}
A variety of direct and iterative solvers are available to solve the system~\eqref{eq:linear_system_FE}, exploiting its regularity, symmetry, and sparsity, e.g.~\cite[Chapter~8]{Bathe:1996:FEP}.

The comparison of~\eqref{eq:nodal_equil_FE} with~\eqref{eq:nodal_equilibrium_equations} reveals that the resulting physical meaning is the same -- i.e. they represent the \emph{nodal equilibrium equations} -- but the expressions differ because of the different parameterizations of the solution. In the FE method, the relation between the nodal unknowns $\M{u}\flc$ and the stresses $\Tstress$ is more involved, because the displacements need to be converted first to strains at the Gauss points via multiplication by the $\M{B}$ matrix. The same holds for the equilibrium conditions, for which the stresses at the Gauss points must be mapped back to the nodal forces by $\M{B}\trn$ with a different weight $\gpw^g$ assigned to each integration point. In the FFT-based method, no exchange of data between the nodes and integration points is needed, because the unknowns $\Mstrainfl$ correspond to strains, and the integration points and nodes coincide. Equilibrium is enforced by the projection matrix $\MG$ with a simple structure inherited from the continuous formulation, see Eq.~\eqref{eq:projection} for the adopted approximation space.

The comparison of the two \emph{linearized systems}~\eqref{eq:linear_system_FE} and~\eqref{eq:linear_system} suggests how to exploit the constitutive routines available in FE systems with FFT-based solvers. Indeed, the only material law-dependent components in~\eqref{eq:linear_system} are the local stresses and the constitutive tangents at the nodes, which can be easily obtained from the FE formulation~\eqref{eq:linear_system_FE}, where the same operation is performed at the Gauss points. In addition, the condition number of the linear system~\eqref{eq:linear_system} depends only on the local consistent constitutive tangents, as discussed next, whereas the conditioning of~\eqref{eq:linear_system_FE} also depends on the mesh size and shape, when unstructured meshes are used, e.g.,~\cite{Ern2006}.

\subsection{Collocation FFT schemes}\label{sec:collocation}

Another reason to adopt the FE recipe when deriving FFT-based methods is to clarify the role of the reference problem used in the conventional approach. The purpose of the current section is to show that the reference problem is an intrinsic choice within the solution algorithm, for which more efficient choices can be made accordingly.

To this purpose, consider a non-linear version of the basic Moulinec-Suquet scheme, e.g.,~\cite[Eq.~(9)]{Moulinec:1998:NMC}, which is based of an integral equation for the fluctuating strains $\Tstrainfl \in \sE$,
\begin{align}\label{eq:L-S}
\int_\puc
\TGamma\aux( \Tx - \Ty )
:
\Tstress\left( \Ty, \TE + \Tstrainfl(\Ty) \right)
\de \Ty
=
\T{0}
\text{ for all }
\Tx \in \puc,
\end{align}
where $\TGamma\aux$ is the \emph{Green function} of the \emph{reference problem} --- an auxiliary local problem~\eqref{eq:local_problem} with the homogeneous constitutive relation
\begin{align}\label{eq:C_aux}
\Tstress\bigl(\Tx, \Tstrain( \Tx ) \bigr)
=
\TC\aux
:
\Tstrain( \Tx )
\text{ for }
\Tx \in \puc.
\end{align}
The constant reference stiffness tensor $\TC\aux$, on which the Green operator $\TGamma\aux$ in~\eqref{eq:L-S} depends, see Eq.~\eqref{eq:Gamma_explicit} in \appref{app:operators}, is yet undetermined in the algorithm , which will be commented on later.

The discretization of Eq.~\eqref{eq:L-S} is then performed by the trigonometric collocation method~\cite[Chapter~10]{Saranen:2000:PIP}, in which we expand the solution $\Tstrainfl$ in terms of the trigonometric polynomials, as in~\eqref{eq:approximation_1a}, and enforce the relation~\eqref{eq:L-S} directly at the grid nodes $\gpoint{\Tk}$ (no numerical quadrature is thus used). As a result, we obtain the following system of non-linear equations for the nodal strains~$\Mstrainfl$, cf.~\eqref{eq:nodal_equilibrium_equations},
\begin{align}\label{eq:L-S_discrete}
\MGamma\aux \,
\Mstress(\ME + \Mstrainfl)
=
\M{0}
\text{ with }
\MGamma\aux = \MF\inv\FT{\MGamma}\aux \MF,
\end{align}
where the matrix $\FT{\MGamma}\aux$ is block-diagonal in the Fourier space; see~\cite{zeman_accelerating_2010} for a more detailed explanation and \appref{app:matrices} for the matrix representations. The remaining steps in the solution of the non-linear system~\eqref{eq:L-S_discrete} now closely follow those of Sections~\ref{sec:linearization} and~\ref{sec:algorithm}, once the projection matrix $\MG$ is replaced with $\MGamma\aux$, including the fact that $\MGamma\aux$ enforces nodal equilibrium and strain compatibility.

Finally, the reference stiffness tensor $\TC\aux$ has to be specified, which was so far done on the basis of local elastic properties~\cite{Moulinec:1998:NMC,Gelebart:2013:NLE}, or of the initial constitutive tangents, Eq.~\eqref{eq:local_tangent} with $i=0$, see~\cite{vinogradov_accelerated_2008}. However, as follows from our developments in Section~\ref{sec:variational} and also from the discussion in~\cite[Section~3]{Mishra:2015:ACL}, this choice rather depends on the \emph{iterative algorithm} used to solve the following linearized system for $\delta\Mstrainfl\ite{\nite + 1}$,
\begin{align}\label{eq:linear_system_L-S}
\MGamma\aux
\MD\ite{\nite}
\delta\Mstrainfl\ite{\nite + 1}
=
- \MGamma\aux
\Mstress(\ME + \Mstrainfl\ite{i}).
\end{align}
For the collocation method this equation replaces~\eqref{eq:linear_system}, which did not depend on a reference medium.

The \emph{basic scheme} from~\cite{Moulinec:1998:NMC} is recovered by solving the system~\eqref{eq:linear_system_L-S} by the Richardson fixed-point iterative method, e.g.,~\cite[Section~3.1]{Mishra:2015:ACL}, which is only conditionally convergent, depending on the choice of $\TC\aux$. Specifically, the optimal convergence is ensured by setting
\begin{align}
\TC\aux =
\frac{1}{2}
\left(
\lambda^\mathrm{min}\ite{\nite}
+
\lambda^\mathrm{max}\ite{\nite}
\right)
\TIs,
\end{align}
where $\TIs$ is the fourth-order symmetric unit tensor. The maximum and minimum eigenvalues, $\lambda^\mathrm{min}\ite{\nite}$ and $\lambda^\mathrm{max}\ite{\nite}$, are defined as

\begin{align}
\lambda^\mathrm{min}\ite{i}
=
\min_{\Tk \in \xZNd}
\lambda_{\min}\left( \MD\ite{\nite}^{\Tk} \right),
&&
\lambda^\mathrm{max}\ite{i}
=
\max_{\Tk \in \xZNd}
\lambda_{\max}\left( \MD\ite{\nite}^{\Tk} \right).
\end{align}
The reference medium thus \emph{must be updated} during the Newton iterations to ensure convergence. For this choice, the number of iterations to reach the given tolerance, $\eta^\mathrm{CG}$ in Algorithm~\ref{fig:algorithm}, grows linearly with the condition number $\lambda^\mathrm{max}\ite{i} / \lambda^\mathrm{min}\ite{i}$.

On the other hand, when the linear system~\eqref{eq:linear_system_L-S} is solved with, e.g., \emph{Conjugate Gradients} as proposed by Zeman\etal~\cite{zeman_accelerating_2010} for linear problems and by Gélébart and~Mondon-Cancel~\cite{Gelebart:2013:NLE} for non-linear problems, it suffices for the convergence of the algorithm that the condition number is \emph{finite}. This in turn implies that the CG method works for \emph{any} choice of reference media, \emph{no updates} of $\TC\aux$ during the Newton increment are needed, and the number of iterations to reach the given accuracy $\eta^\mathrm{CG}$ grows as $\sqrt{\lambda^\mathrm{max}\ite{i} / \lambda^\mathrm{min}\ite{i}}$, cf.~\cite[Section 5.1]{vondrejc_fft-based_2014} or ~\cite[Section~3.2]{Mishra:2015:ACL}. Therefore, the simplest option is to take $\TC\aux = \TIs$, for which $\TGamma\aux = \TG$, see \appref{app:operators}, whereby the collocation and the variational formulations coincide.

Even though the collocation and variational approaches become equivalent for specific choices of the reference stiffness tensor $\TC\aux$ and the iterative solver, the variational formulation offers at least two advantages. First, it clarifies the connection between the strain compatibility, equilibrium conditions, and the reference problem in non-linear homogenization, which has been a source of confusion in the FFT-based literature. Second, it enables us to interpret and understand the Fourier-based technique in the language of (spectral) FE methods, so that the extensive knowledge accumulated in the field of non-linear Finite Elements may be explored when developing Fourier solvers beyond small-strain computational inelasticity.

For the reader's convenience, we conclude this section by summarizing the most important characteristics in \tabref{tab:comparison}.

\begin{table}[h]
\centering
\caption{Comparison of FFT-based and Finite Element methods.}
\label{tab:comparison}
\small
\begin{tabular}{lccc}
\hline
& {\bf Finite elements}      & {\bf Conventional FFT} & {\bf Variational FFT} \\
\hline
Discretization approach      & Galerkin         & collocation    & Galerkin         \\
Computational grid           & general          & regular        & regular          \\
Basis functions              & Lagrange         & trigonometric  & trigonometric    \\
Unknown                      & displacement     & strain         & strain           \\
Compatibility of solution    & automatic        & linear solver  & linear solver    \\
Compatibility of test fields & automatic        & $\times$       & projection matrix $\MG$ \\
Equilibrium                  & static matrix $\M{B}\trn$ & Green matrix $\MGamma\aux$ & projection matrix $\MG$ \\
Reference problem       & $\times$         & yes            & $\times$      \\
Quadrature              & Gauss            & $\times$       & trapezoidal   \\
Linear system & regular symmetric, & singular non-symmetric, & singular non-symmetric, \\
& sparse & structurally sparse & structurally sparse
\\
Linear system solver    & direct/iterative & iterative      & iterative     \\
\hline
\end{tabular}
\end{table}

\section{Constitutive models and their numerical implementation}
\label{sec:model}

As pointed out above, the proposed FFT scheme is general and robust in the sense that arbitrary constitutive models formulated in small strain framework may be inserted at the integration point level. To demonstrate this feature, we consider three different constitutive models -- non-linear elasticity, elasto-plasticity, and visco-plasticity -- which are non-linear or/and history dependent. Each of these models is discussed briefly below, together with its numerical treatment. More details for the elasto- and visco-plastic models can be found in textbooks, e.g. \cite{simo_computational_1998,neto_computational_2011}. Note that the same symbols are used in the different models, their exact meaning and quantification may however be different.

\subsection{Non-linear elasticity}
\label{sec:model:nlin}

\paragraph{Model}

The following non-linear elastic model is considered:
\begin{equation}\label{eq:model:nlin:stress}
  \bm{\sigma}
  = K \mathrm{tr}\, ( \bm{\varepsilon} ) \, \bm{I}
  + \sigma_0 \left( \frac{\varepsilon_\mathrm{eq}}{\varepsilon_0} \right)^n \,
    \bm{N},
\end{equation}
where $\bm{I}$ is the second-order identity tensor, $\bm{N}$ is now the direction of the deviatoric strain defined as
\begin{equation}\label{eq:model:nlin:N}
  \bm{N} =
  \frac{2}{3} \frac{\bm{\varepsilon}_\mathrm{d}}{\varepsilon_\mathrm{eq}};
\end{equation}
and the equivalent strain, $\varepsilon_\mathrm{eq}$, is defined as
\begin{equation}
  \varepsilon_\mathrm{eq}
  = \; \sqrt{
    \tfrac{2}{3} \, \bm{\varepsilon}_\mathrm{d} : \bm{\varepsilon}_\mathrm{d}
  },
\end{equation}
with $\bm{\varepsilon}_\mathrm{d}$ the strain deviator. The parameters are the bulk modulus $K$, a reference shear stress $\sigma_0$ and strain $\varepsilon_0$, and an exponent $n$.

\paragraph{Stress update}

Since this model does not depend on the deformation history, the stress can directly be evaluated from Eq.~\eqref{eq:model:nlin:stress} for every increment.

\paragraph{Consistent constitutive tangent}

The consistent tangent operator is obtained by taking the derivative of \eqref{eq:model:nlin:stress} with respect to the strain $\bm{\varepsilon}$, i.e.:
\begin{equation}\label{eq:model:nlin-tangent}
  \T{C} =
  \frac{\partial \bm{\sigma}}{\partial \bm{\varepsilon}} =
  K \bm{I} \otimes \bm{I} +
  \frac{\sigma_0}{\varepsilon_\mathrm{eq}}
  \left( \frac{\varepsilon_\mathrm{eq}}{\varepsilon_0} \right)^n
  \Big(
    (n-1) \, \bm{N} \otimes \bm{N}
    + \tfrac{2}{3} \, \bm{I}_\mathrm{d}
  \Big),
\end{equation}
with $\T{I}_\mathrm{d} = \TIs - \tfrac{1}{3} \T{I} \otimes \T{I}$ the fourth order deviatoric identity tensor.

\subsection{Elasto-plasticity}
\label{sec:model:plas}

\paragraph{Model}

Standard $J_2$-plasticity is considered. In this model the total strain, $\T{\varepsilon}$, is additively split into an elastic part, $\T{\varepsilon}_\mathrm{e}$, and a plastic part, $\bm{\varepsilon}_\mathrm{p}$, i.e.\
\begin{equation}\label{eq:model:plas:split}
  \T{\varepsilon} =
  \T{\varepsilon}_\mathrm{e} + \T{\varepsilon}_\mathrm{p}.
\end{equation}

The stress, $\bm{\sigma}$, depends on the elastic strain, $\bm{\varepsilon}_\mathrm{e}$, through the standard linear relation:
\begin{equation}\label{eq:model:plas:stress}
  \bm{\sigma}
  = \bm{C}_\mathrm{e} :
  \big( \bm{\varepsilon} - \bm{\varepsilon}_\mathrm{p} \big),
  \qquad\text{with}\qquad
  \bm{C}_\mathrm{e}
  = K \bm{I} \otimes \bm{I}
  + 2 G \, \bm{I}_\mathrm{d},
\end{equation}
wherein $K$ is the bulk modulus and $G$ is the shear modulus.

The elastic domain is bounded by the plastic admissibility condition
\begin{equation}\label{eq:model:plas:yield}
  \Phi( \bm{\sigma} , \varepsilon_\mathrm{p} )
  = \sigma_\mathrm{eq}
  - ( \sigma_0 + H \varepsilon_\mathrm{p}^n ) \leq 0,
\end{equation}
wherein the parameters are the initial yield stress, $\sigma_0$, the hardening modulus, $H$, and the hardening exponent, $n$. The deformation history enters this expression via the accumulated plastic strain, $\varepsilon_\mathrm{p}$ (which equals zero in the initial stress-free state). Finally, the von Mises equivalent stress, $\sigma_\mathrm{eq}$, is defined as
\begin{equation}\label{eq:model:eq-stress}
  \sigma_\mathrm{eq} =
  \sqrt{\tfrac{3}{2} \bm{\sigma}_\mathrm{d} : \bm{\sigma}_\mathrm{d}},
\end{equation}
with $\bm{\sigma}_\mathrm{d}$ the stress deviator.

The plastic strain rate follows from normality as
\begin{equation}\label{eq:model:plas:flow}
  \dot{\bm{\varepsilon}}_\mathrm{p}
  = \dot{\gamma} \bm{N}
  = \dot{\gamma}\, \frac{\partial \Phi}{\partial \bm{\sigma}}
  = \dot{\gamma}\, \frac{3}{2}\frac{\bm{\sigma}_\mathrm{d}}{\sigma_\mathrm{eq}},
\end{equation}
where $\bm{N}$ is defined differently from \eqref{eq:model:nlin:N}. The accumulated plastic strain is determined from
\begin{equation}\label{eq:model:plas:eq-plas}
  \varepsilon_\mathrm{p} =
  \int\limits_0^t \dot{\varepsilon}_\mathrm{p} ~\mathrm{d}t^\prime,
  \qquad\text{with}\qquad
  \dot{\varepsilon}_\mathrm{p}
  = \sqrt{
      \tfrac{2}{3} \,
      \dot{\bm{\varepsilon}}_\mathrm{p} : \dot{\bm{\varepsilon}}_\mathrm{p}
    }
  = \dot{\gamma}.
\end{equation}
The reader is reminded that in this model the time-derivative is used just for convenience, the model is completely rate-independent.

\paragraph{Stress update}

The model is discretized in time using the, unconditionally stable, backward Euler scheme. The stress update is implemented using an elastic-predictor plastic-corrector scheme, whereby the amount of plastic flow is determined in two steps. First, a trial state is calculated by assuming the increment in strain to be fully elastic (\textit{elastic predictor}). Second, if necessary, a return-map is used that quantifies the plastic strain increment (\textit{plastic corrector}).

Given an increment in total strain
\begin{equation}\label{eq:model:plas:strain-inc}
  \Delta \bm{\varepsilon}
  = \bm{\varepsilon}^{(t+\Delta t)} - \bm{\varepsilon}^{(t)}
\end{equation}
(where $\Delta t$ refers to a pseudo-time step), the trial state (\textit{elastic predictor}, denoted by $\leftstar \bullet$) is computed by assuming that $\Delta \bm{\varepsilon}$ gives rise to a purely elastic strain increment, i.e.:
\begin{equation}\label{eq:model:plas:trial}
  \leftstar{\bm{\varepsilon}}_\mathrm{p}
  = \bm{\varepsilon}_\mathrm{p}^{(t)}
  \qquad\text{and}\qquad
  \leftstar{\varepsilon}_\mathrm{p}
  = \varepsilon_\mathrm{p}^{(t)}.
\end{equation}
The trial stress, $\leftstar{\bm{\sigma}}$, is found by evaluating Eq.~\eqref{eq:model:plas:stress} , using $\Tstrain = \Tstrain^{(t+\Delta t)}$ and $\Tstrain_\mathrm{p}=\leftstar{\bm{\varepsilon}}_\mathrm{p}$.

The yield function in Eq.~\eqref{eq:model:plas:yield} can now be evaluated for the trial stress $\leftstar{\bm{\sigma}}$. If $\leftstar{\Phi} \leq 0$, the current increment does not give rise to plastic flow. The actual state thus coincides with the trial state, and
\begin{equation}\label{eq:model:plas:elas}
  \bm{\varepsilon}_\mathrm{p}^{(t + \Delta t)} =
  \leftstar{\bm{\varepsilon}}_\mathrm{p} =
  \bm{\varepsilon}_\mathrm{p}^{(t)},
  \quad
  \varepsilon_\mathrm{p}^{(t + \Delta t)} =
  \leftstar{\varepsilon}_\mathrm{p} = \varepsilon_\mathrm{p}^{(t)},
  \quad
  \bm{\sigma}^{(t + \Delta t)} =
  \bm{C}_\mathrm{e} :
  \big( \bm{\varepsilon}^{(t + \Delta t)} - \bm{\varepsilon}_\mathrm{p}^{(t )} \big).
\end{equation}

If $\leftstar{\Phi} > 0$, a return-map (\textit{plastic corrector}) has to be performed to return the trial state to an admissible state. For this state, the equality needs to hold in the yield function (Eq.~\eqref{eq:model:plas:yield}), given the actual stress that in turn depends on the plastic flow (Eqs.~(\ref{eq:model:plas:flow},~\ref{eq:model:plas:eq-plas})). Due to the assumed normality (Eq.~\eqref{eq:model:plas:flow}), this non-linear system of equations can be rewritten as a single scalar equation:
\begin{equation}
  \Phi
  = \leftstar{\sigma}_\mathrm{eq}
  - 3 G \Delta \gamma
  - \sigma_0 - H ( \, \varepsilon_\mathrm{p}^{(t)} + \Delta \gamma \, )^n
  = 0,
\end{equation}
which has to be solved for $\Delta \gamma$ (in closed form for $n = 1$, or numerically for arbitrary $n$). The resulting state can is then determined as
\begin{equation}\label{eq:model:plas:corrector}
  \bm{\varepsilon}_\mathrm{p}^{(t + \Delta t)} =
  \bm{\varepsilon}_\mathrm{p}^{(t)} + \Delta \gamma \;\leftstar{\bm{N}},
  \quad
  \varepsilon_\mathrm{p}^{(t + \Delta t)} =
  \varepsilon_\mathrm{p}^{(t)} + \Delta \gamma,
  \quad
  \bm{\sigma}^{(t + \Delta t)} =
  \bm{C}_\mathrm{e} :
  \big( \bm{\varepsilon}^{(t + \Delta t)} - \bm{\varepsilon}_\mathrm{p}^{(t + \Delta t)} \big).
\end{equation}
%

\paragraph{Consistent constitutive tangent}

The tangent is easily derived by linearizing the stress update procedure. If the trial state is elastic, i.e.\ when $\leftstar{\Phi} \leq 0$, the result is trivially $\bm{C} = \bm{C}_\mathrm{e}$. Otherwise, the stress update in Eq.~\eqref{eq:model:plas:corrector} needs to be linearized, giving
\begin{align}
  \bm{C} & =
  \frac{
    \partial \bm{\sigma}^{(t+\Delta t)}
  }{
    \partial \bm{\varepsilon}^{(t+\Delta t)}
  }
  \nonumber \\
  & =
  \bm{C}_\mathrm{e} -
  \frac{6 G^2 \Delta \gamma}{\leftstar{\sigma}_\mathrm{eq}}
  \bm{I}_\mathrm{d}
  + 4 G^2
  \left(
    \frac{\Delta \gamma}{\leftstar{\sigma}_\mathrm{eq}} -
    \frac{1}{3 G + n H \big( \varepsilon_\mathrm{p}^{(t)} + \Delta \gamma \big)^{n-1}}
  \right)
  \leftstar{\bm{N}} \otimes \leftstar{\bm{N}}.
\end{align}
%

\subsection{Visco-plasticity}
\label{sec:model:visco}

\paragraph{Model}

The considered visco-plastic model has many similarities to the elasto-plastic model of the previous section. The only differences are that the visco-plastic model is rate-dependent and that there is no discrete switch between elasticity and plasticity (i.e.\ there is plastic flow at each stage of deformation). Similar to elasto-plasticity, the model is governed by an additive split of elastic and plastic strains (Eq.~\eqref{eq:model:plas:split}). The stress can be expressed by the elastic strain only (Eq.~\eqref{eq:model:plas:stress}). The direction of plastic flow, $ \dot{\bm{\varepsilon}}_\mathrm{p}
  = \dot{\gamma} \bm{N},
$ is determined similarly as in Eq.~\eqref{eq:model:plas:flow},  however the plastic rate depends on the stress through Norton's rule
\begin{equation}\label{eq:model:visco:gammadot}
  \dot{\gamma}
  =
  \frac{\varepsilon_0}{t_0}
  \left( \frac{\sigma_\mathrm{eq}}{\sigma_0} \right)^{1/n}.
\end{equation}
In this equation $\varepsilon_0$, $t_0$, $\sigma_0$, and $n$ are material parameters. Note that $n$ in this case is the strain rate sensitivity exponent, which has a different meaning than $n$ in the elasto-plastic model.

\paragraph{Stress update}

A backward Euler scheme is used for discretization in time. Even though the actual physical process is never elastic, a trial state in conjunction with a return-map is again employed. This has the benefit that the plastic strain can be determined by solving a single scalar equation. In particular, given an increment in strain (Eq.~\eqref{eq:model:plas:strain-inc}), the trial state (\textit{elastic predictor}) is given by Eq.~\eqref{eq:model:plas:trial}, where $\Delta t$ now refers to a real time step. A \textit{plastic corrector} is needed to enforce \eqref{eq:model:visco:gammadot}, leading to the following implicit equation for $\Delta \gamma$:
\begin{equation}
  \Delta \gamma
  =
  \frac{\varepsilon_0}{t_0} \Delta t \left(
    \frac{
      \leftstar{\sigma_\mathrm{eq}} - 3 G \Delta \gamma
    }{
      \sigma_0
    }
  \right)^{1/n},
\end{equation}
which is solved numerically. The plastic strain and stress are then determined from Eq.~\eqref{eq:model:plas:corrector}.

\paragraph{Consistent constitutive tangent}

The consistent tangent is obtained again by linearizing the stress update. The result reads
\begin{equation}
  \bm{C}
  =
  \bm{C}_\mathrm{e} -
  \frac{6 G^2 \Delta \gamma}{\leftstar{\sigma}_\mathrm{eq}} \bm{I}_\mathrm{d}
  + 4 G^2
  \left(
    \frac{\Delta \gamma}{\leftstar{\sigma}_\mathrm{eq}} -
    \left(
      3G + \frac{n \sigma_0}{\gamma_0 \Delta t}
      \left(
        \frac{\Delta \gamma}{\gamma_0 \Delta t}
      \right)^{n-1}
    \right)^{-1}
  \right)
  \leftstar{\bm{N}} \otimes \leftstar{\bm{N}}.
\end{equation}
%

\section{Examples}
\label{sec:Examples}

\subsection{Two-phase laminate}

\begin{figure}[ht]
  \centering
  \includegraphics[width=.5\textwidth]{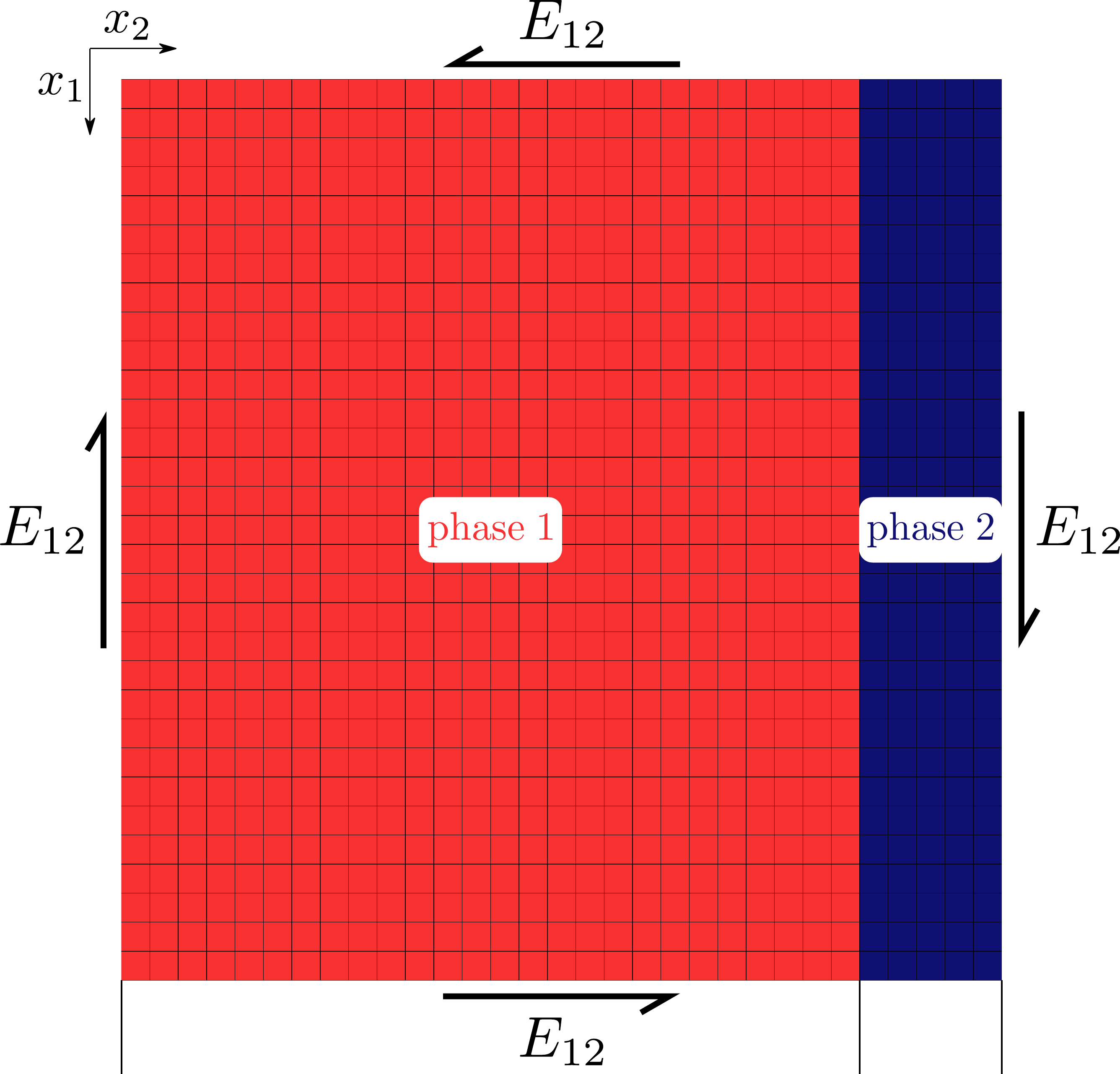}
  \caption{Two-phase laminate. Phase 1 is modeled using different materials models: non-linear elastic, elasto-plastic, and visco-plastic; phase 2 is always linear elastic. The applied shear is indicated by $E_{12}$.}
  \label{fig:ex-laminate_geometry}
\end{figure}

The goal of this section is to demonstrate the accuracy and the convergence rate of the Newton-based FFT algorithm. We consider a periodic two-phase laminate subjected to shear, see Figure~\ref{fig:ex-laminate_geometry}. In this figure, the numerical discretization is also indicated, whereby each pixel corresponds to one grid point in its center. The applied global shear is indicated by arrows, corresponding to a global strain tensor
\begin{equation}\label{eq:shear_strain}
  \TE =
  E_{12}
  \left(
  \T{e}_1 \otimes \T{e}_2 + \T{e}_2 \otimes \T{e}_1
  \right),
\end{equation}
wherein $E_{12}$ is the global shear strain. Phase 1 is modeled with different material models: non-linear elastic, elasto-plastic, and visco-plastic. Phase 2 is taken to be linear elastic in all cases. The elastic properties of the two phases are identical, except for the non-linear elastic case.

The used material parameters are listed in Table~\ref{tab:ex-laminate_parameters}. The total overall shear strain is set to $E_{12} = 0.05$. Note that for the non-linear elastic and the elasto-plastic model a single time increment suffices to obtain the exact solution. For the visco-plastic model the deformation is applied in $200$ equi-sized increments, each with a time step of $10^{-3}$ seconds. The tolerances are set to $\eta^\text{NW} = 10^{-6}$ for the Newton iterations and $\eta^\text{CG} = 10^{-16}$ for the Conjugate Gradient iterative solver.

\begin{table}[ht]
  \centering
  \caption{The material parameters of phase 1 for the different material models. Phase 2 is linear elastic with shear modulus $G$.}
  \label{tab:ex-laminate_parameters}
  \begin{tabular}{c||c|c|c}
  parameter       & non-linear elasticity & elasto-plasticity & visco-plasticity
  \\ \hline \hline
  $H/G$           & $-$                   & $0.05$            & $-$           \\
  $\sigma_0/G$    & $0.5$                 & $0.01$            & $0.1$         \\
  $\varepsilon_0$ & $0.1$                 & $-$               & $0.1$         \\
  $t_0$           & $-$                   & $-$               & $0.1$         \\
  $n$             & $10.0$                & $0.1$             & $0.3$         \\
  \end{tabular}
\end{table}

\begin{figure}[htp]
  \centering
  \includegraphics[width=1.\textwidth]{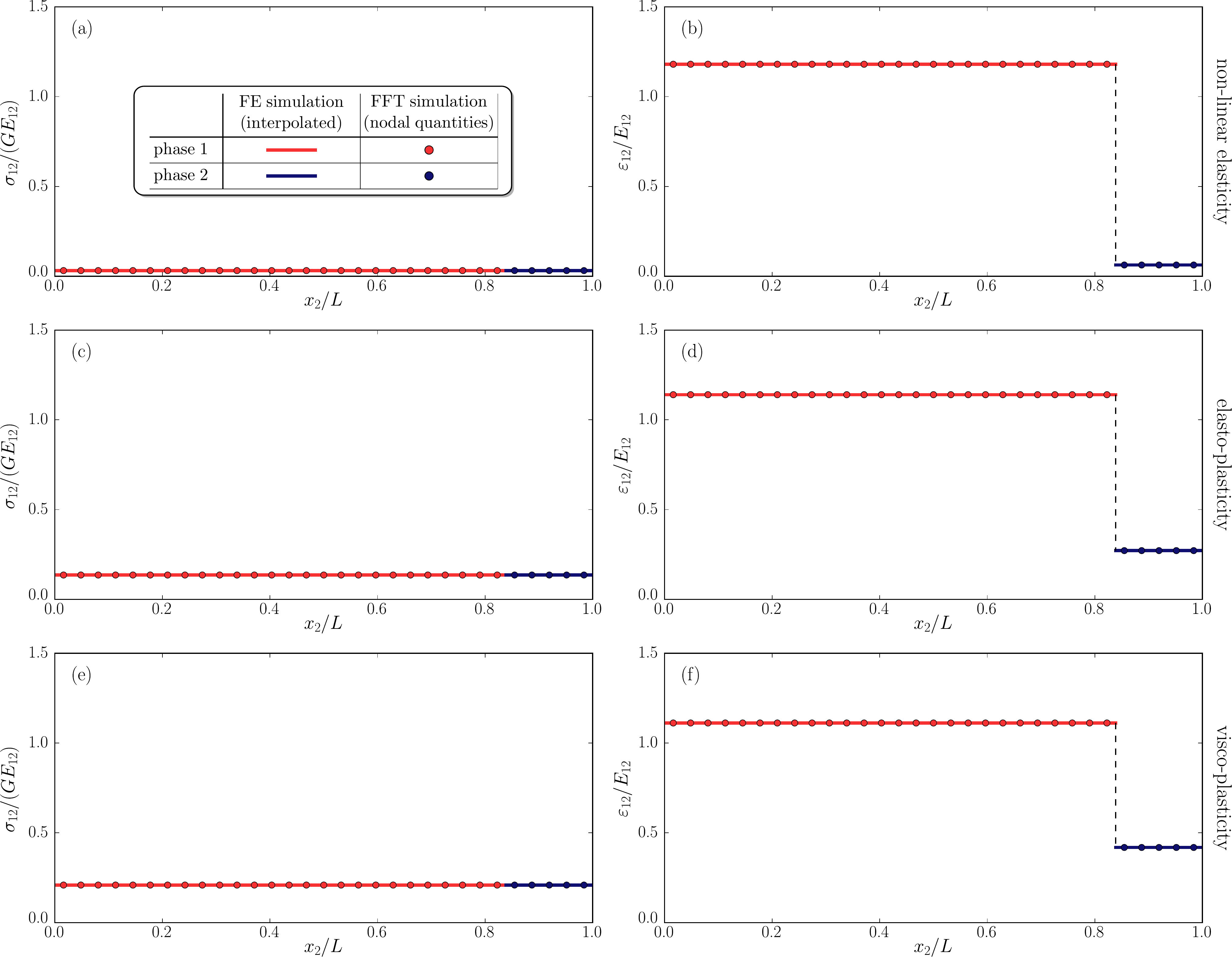}
  \caption{The shear stress $\sigma_{12}$ (left) and the shear strain $\varepsilon_{12}$ (right), both along the $x_2$-direction (the response does not depend on $x_1$). From top to bottom the different material models for phase 1: (a--b) non-linear elasticity, (c--d) elasto-plasticity, and (e--f) visco-plasticity. The predicted numerical response is shown using a marker at each node / integration point and the result of a FE simulation of two elements using solid lines; for both, the color corresponds to the phase (cf.\ Figure~\ref{fig:ex-laminate_geometry}). The stress is normalized by the shear modulus, $G$, and a reference elastic stress that accompanies the applied strain. The strain is normalized by the applied shear strain, $E_{12}$.}
  \label{fig:ex-laminate_response}
\end{figure}

The results are presented in Figure~\ref{fig:ex-laminate_response}. All diagrams in this figure are cross-sections of the corresponding fields along the $x_2$-axis whereby the red and blue color correspond to phase 1 and 2 respectively, cf.\ Figure~\ref{fig:ex-laminate_geometry}. The response is constant in $x_1$-direction. The numerical response is included using a marker for each node / integration point. In each case, we show a comparison with the response of a FE simulation (solid lines) of just two elements (one per phase), which for this case resolves the problem \textit{exactly} in space. The rows correspond to the different considered material models for phase 1; the left column shows the distribution of shear stress $\sigma_{12}$, the right column shows the shear strain $\varepsilon_{12}$. The results reveal a perfect agreement: a constant shear stress $\sigma_{12}$ and a piece-wise constant shear strain $\varepsilon_{12}$. A perfect agreement is also found when the responses of the simulations are compared to analytical solutions, for which the exponents are set to $n = 1$ (results not shown).

\begin{figure}[htp]
  \centering
  \includegraphics[width=.5\textwidth]{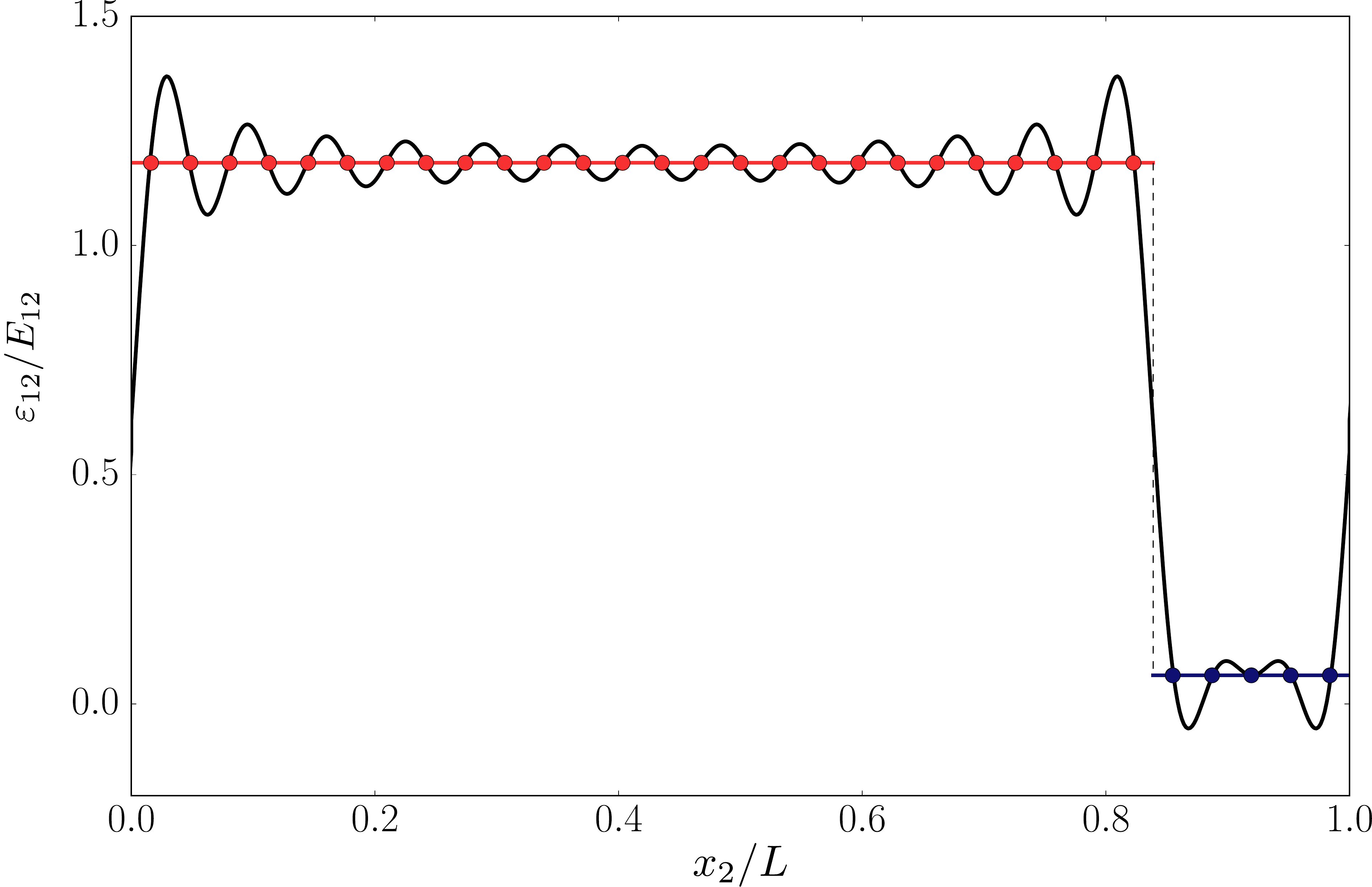}
  \caption{The interpolation of the nodal response using the trigonometric polynomials, according to~\eqref{eq:approximation_1a}, for the non-linear elastic model (cf.\ Figure~\ref{fig:ex-laminate_response}(b)). The interpolation is shown using a solid black line, in addition to nodal quantities (markers) and the FE result (solid red and blue lines).}
  \label{fig:ex-laminate_response-interp}
\end{figure}

It is also observed from Figure~\ref{fig:ex-laminate_response} that while the trigonometric interpolation may not be able to fully capture the step in the response because of Gibbs phenomena, at the nodes / integration points no artifacts occur. When the nodal quantities are interpolated using the trigonometric basis functions, such oscillations are however clearly observed: see the solid black line in Figure~\ref{fig:ex-laminate_response-interp}. This is in agreement with validation studies by Moulinec and Suquet~\cite{Moulinec:1998:NMC} and Anglin\etal~\cite{Anglin:2014:VOM}, where a good match with analytical solutions at the \emph{grid points} has been reported for several elastic benchmarks.

To verify that the convergence is quadratic, the residual at the end of each iteration is listed in Table~\ref{tab:ex-laminate_convergence}. In all cases, the quadratic convergence has indeed been achieved, by virtue of the use of consistent tangent operators in the FFT algorithm.

\renewcommand{\arraystretch}{1.15}
\begin{table}[ht]
  \centering
  \caption{The stress residual for each iteration of the Newton process for the different, non-linear, material models.}
  \label{tab:ex-laminate_convergence}
  \begin{tabular}{c||c|c|c}
  iteration & non-linear elasticity & elasto-plasticity & visco-plasticity
  \\ \hline \hline
  1                     &
  $4.23 \cdot 10^{-01}$ &
  $3.19 \cdot 10^{-01}$ &
  $1.33 \cdot 10^{-01}$
  \\
  2                     &
  $1.24 \cdot 10^{-02}$ &
  $1.26 \cdot 10^{-04}$ &
  $7.39 \cdot 10^{-05}$
  \\
  3                     &
  $2.44 \cdot 10^{-05}$ &
  $2.70 \cdot 10^{-10}$ &
  $6.49 \cdot 10^{-10}$
  \\
  4                     &
  $9.18 \cdot 10^{-11}$ &
                        &
  \end{tabular}
\end{table}

\subsection{Application: dual-phase steel}

\begin{figure}[ht]
  \centering
  \includegraphics[width=.7\textwidth]{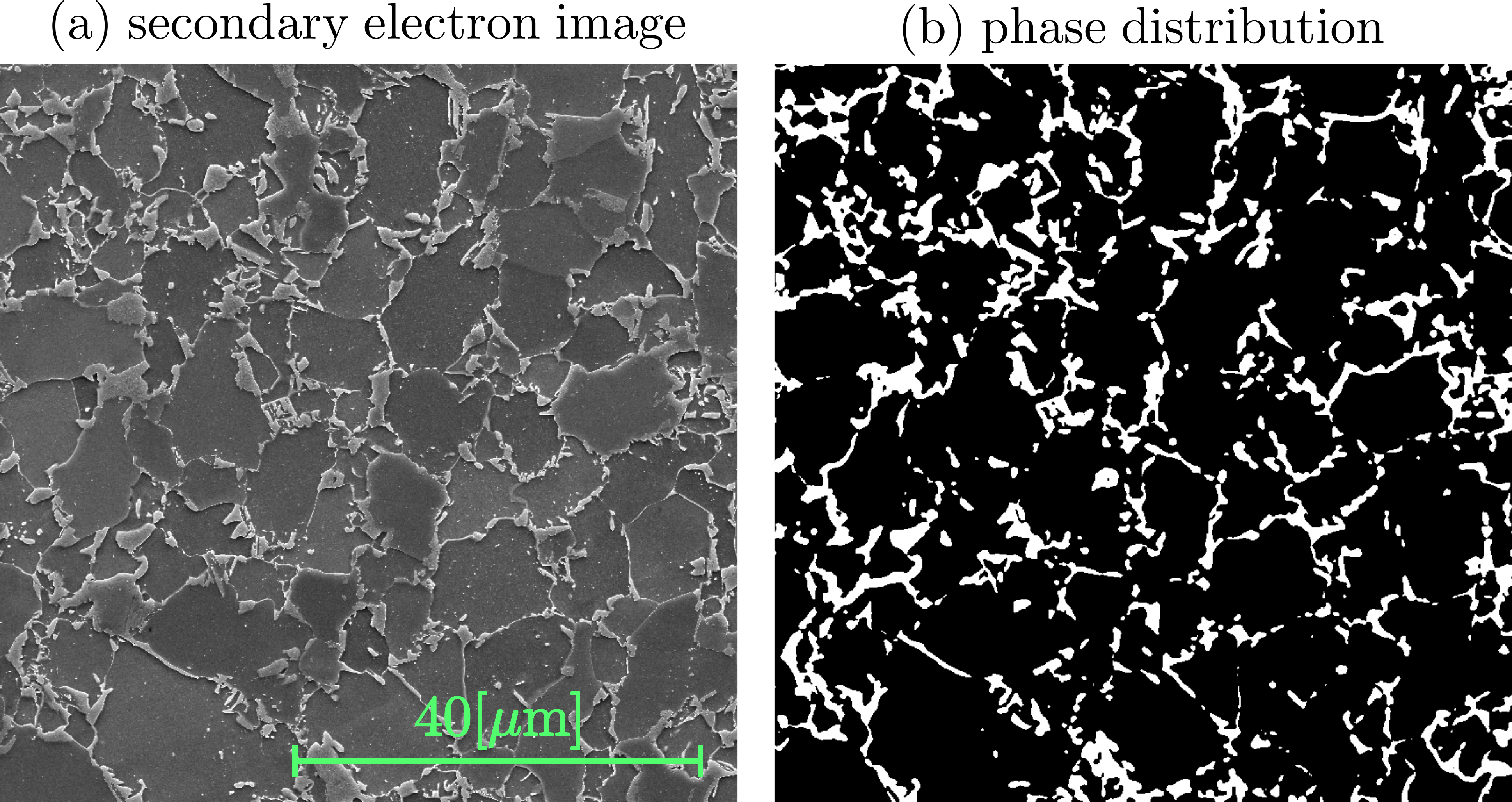}
  \caption{(a) An SEM micrograph of commercial dual-phase steel (DP600) taken in secondary electron mode. (b) The result of the image intensity thresholding: the identified hard martensite is white, the soft ferrite is black.}
  \label{fig:ex-dp_experiment}
\end{figure}

To demonstrate the practical applicability of the method, the microstructural response of a commercial dual-phase steel (DP600) is studied. This steel has a complex microstructure comprising a relatively hard but brittle martensite phase that acts as reinforcement of the comparatively soft yet ductile ferritic matrix phase. Minor fractions of several other phases are frequently observed, however this is disregarded in the present work. To obtain the cell $\puc$, a steel sheet is imaged in the cross-section using a scanning electron microscope. A protocol of grinding, polishing, and etching is applied to create a surface with a small height difference between martensite and ferrite. This provides contrast in the secondary electron mode of a scanning electron microscope~(SEM), as shown in Figure~\ref{fig:ex-dp_experiment}(a). In this image, the bright regions are martensite while the darker regions are ferrite. The phase distribution can be obtained by thresholding, combined with a Gaussian filter to reduce local artifacts due to image noise. The result is shown in Figure~\ref{fig:ex-dp_experiment}(b), for which it is found that the hard phase volume fraction equals $17 \%$.

Both phases are modeled using the isotropic rate-independent elasto-plastic model of Section~\ref{sec:model:plas}. The parameters are taken more or less representative for the martensite phase, denoted ``hard'' below, and the ferrite phase, denoted ``soft'' below. The initial yield stresses and the hardening moduli of the two phases are
\begin{align}
  \frac{\sigma_0^\mathrm{hard}}{E} = 2 \
  \frac{\sigma_0^\mathrm{soft}}{E} =
  1.7 \cdot 10^{-4},
&&
  \frac{H^\mathrm{hard}}{E} = 2 \
  \frac{H^\mathrm{soft}}{E} =
  2.6 \cdot 10^{-4},
\end{align}
and the hardening exponent is set to
\begin{equation}
  n^\mathrm{hard} = n^\mathrm{soft} = 0.2
\end{equation}
The elastic properties are identical for both phases, with the Poisson ratio $\nu = 0.3$.

A macroscopic pure shear deformation
\begin{equation}
  \TE =
  \frac{\sqrt{3}}{2}\, E_\mathrm{eq} \;
  \big(
    \T{e}_2 \otimes \T{e}_2 - \T{e}_1 \otimes \T{e}_1
  \big)
\end{equation}
is applied to this microstructural volume element. The global equivalent strain, $E_\mathrm{eq}$, is imposed in~$200$ equi-sized increments up to the value of~$0.1$. A finite strain assumption would be appropriate for such strain levels, in particular because the magnitude of local strains is further amplified by the microstructural arrangement. Nevertheless,
the purpose of this example is to demonstrate robustness of the solver for highly nonlinear problems, the small strain framework is therefore sufficient.

\begin{figure}[ht]
  \centering
  \includegraphics[width=1.\textwidth]{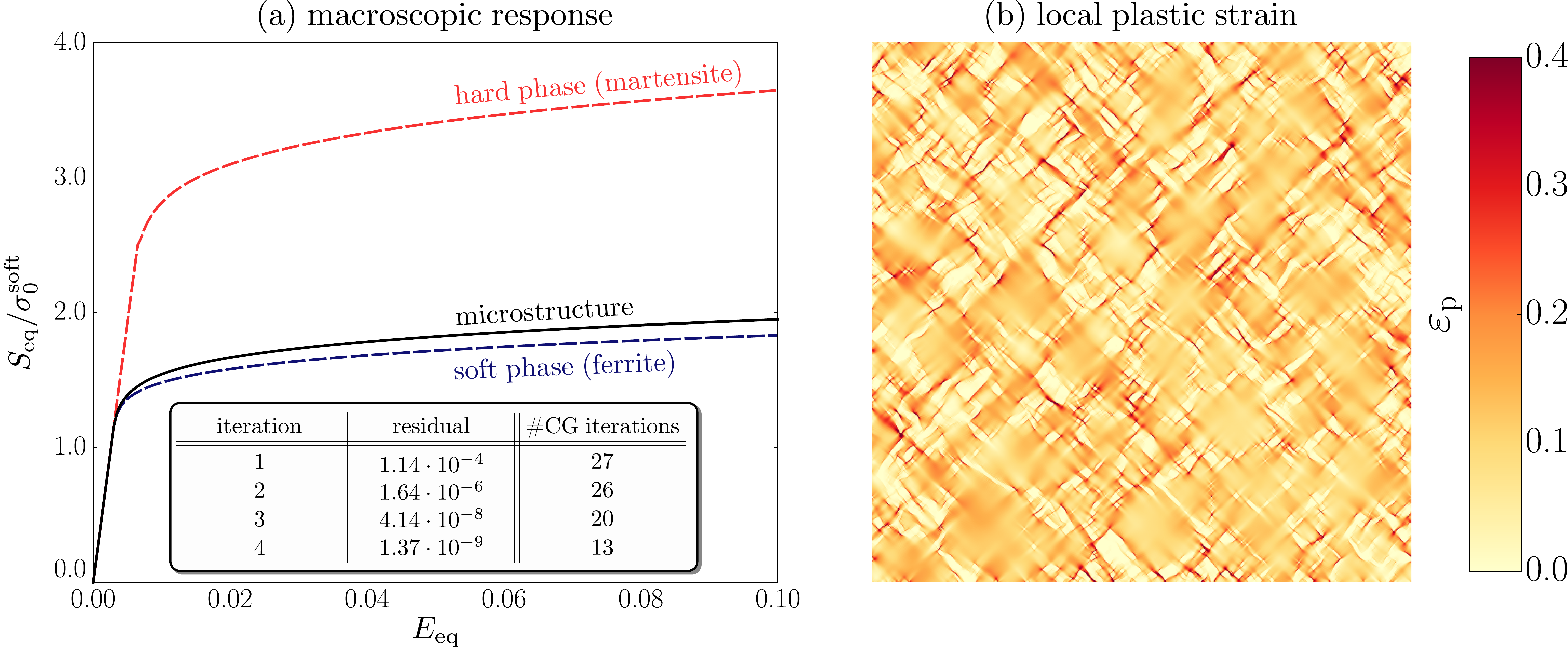}
  \caption{(a) The macroscopic equivalent stress $S_\mathrm{eq}$ as a function of the applied equivalent strain $E_\mathrm{eq}$. The convergence of the Newton iterations and the number of iterations of the conjugate gradient algorithm are indicated for a representative increment ($E_\mathrm{eq} = 0.01$). (b) The local accumulated plastic strain $\varepsilon_\mathrm{p}$ at the final increment of applied strain ($E_\mathrm{eq} = 0.1$). }
  \label{fig:ex-dp_result}
\end{figure}

The macroscopic response is shown in Figure~\ref{fig:ex-dp_result}(a) in terms of the macroscopic equivalent stress $S_\mathrm{eq}$ as a function of the applied equivalent strain $E_\mathrm{eq}$ (solid black line). The constitutive response of the two phases is also included using colored dashed lines. As observed, the predicted response is a non-linear combination of that of its constituting phases phases. In this figure also the convergence is tabulated, revealing that the convergence is  no longer quadratic. This is a well-known limitation for an elasto-plastic model, which is caused by the on/off switch for yielding, accompanied by a significant differences in the tangent stiffness, e.g.~\cite{Blaheta:1997:CNT}. This effect is enhanced by the complex microstructure. Still, the method remains robust as no convergence difficulties were encountered during the simulation.

\begin{figure}[ht]
  \centering
  \includegraphics[width=.8\textwidth]{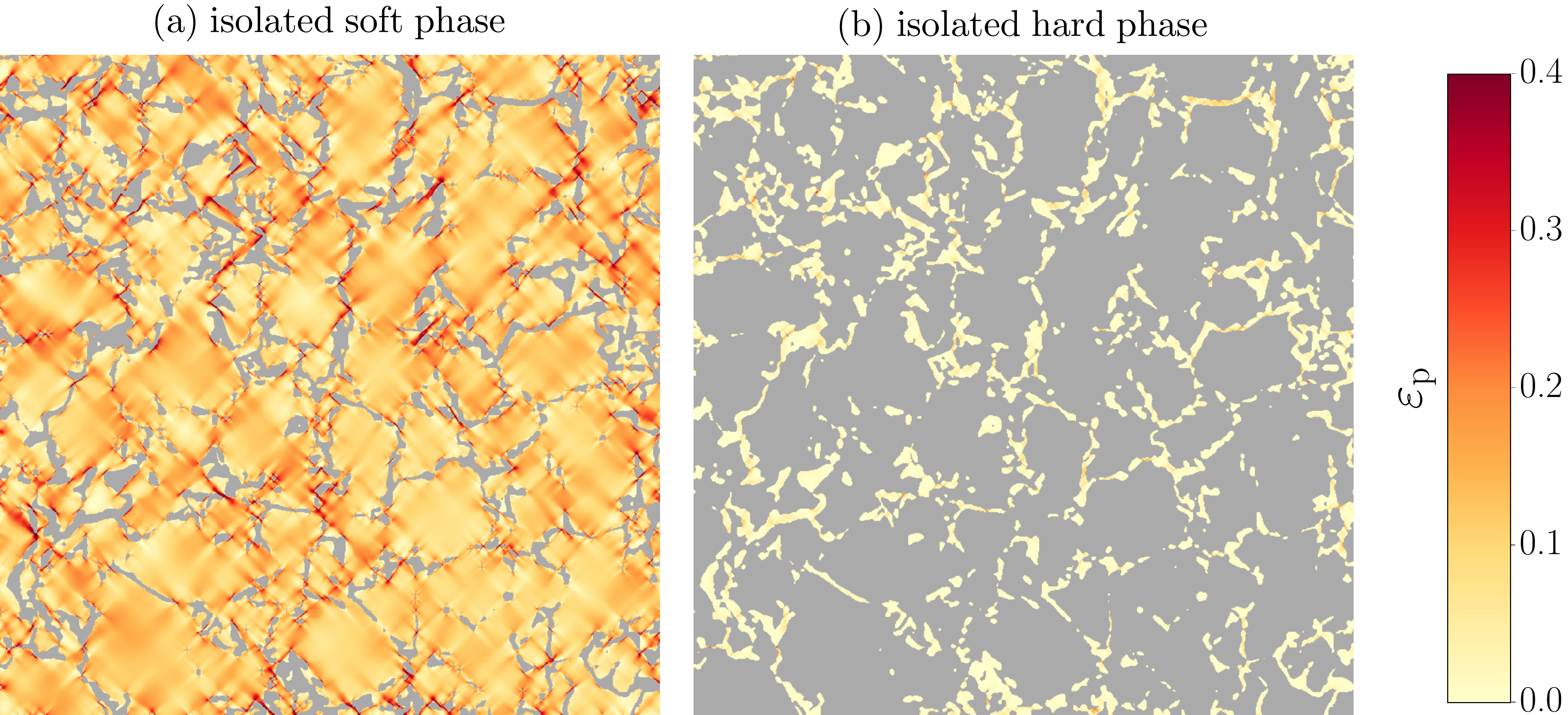}
  \caption{The local equivalent plastic strain $\varepsilon_\mathrm{p}$ at the final increment of applied strain ($E_\mathrm{eq} = 0.1$) for the (a) soft phase and (b) hard phase.}
  \label{fig:ex-dp_result-phases}
\end{figure}

The local response is shown in Figure~\ref{fig:ex-dp_result}(b) in the form of the accumulated plastic strain $\varepsilon_\mathrm{p}$. As observed, the plastic flow is concentrated in bands that are oriented at $\pm 45$ degree angles. These angles correspond to the direction of maximum shear set by the applied macroscopic deformation. The percolation in bands is fully determined by the microstructure. To better understand this, the plastic response is plotted for each phase separately in Figure~\ref{fig:ex-dp_result-phases} revealing that the plastic strain is obviously higher in the soft phase (Figure~\ref{fig:ex-dp_result-phases}(a)) than in the hard phase (Figure~\ref{fig:ex-dp_result-phases}(b)). Furthermore, it is observed that the plastic strain is localized in bands in the soft phase, wherever it is close to the hard phase. This localization pattern is most pronounced where the separation of the islands of the hard phase is small.

\section{Conclusions}
\label{sec:conclusion}

A Fast Fourier Transform~(FFT)-based incremental-iterative solver for micromechanical simulations of heterogeneous media has been developed that can deal with non-linear, history- and time-dependent materials laws under small strains. Contrary to conventional approaches derived from integral equations of the Lippmann-Schwinger type, the proposed formulation aligns the standard procedures used in non-linear Finite Element methods. Specifically, we have (i) discretized the strain-based weak form of the local cell problem with trigonometric polynomials, (ii) approximated the integrals with trapezoidal quadrature, and (iii) solved the resulting system of non-linear nodal equilibrium equations with a Newton scheme that employs consistent linearization to obtain a lineared system, which is solved iteratively with the Conjugate Gradient algorithm. The method has been successfully verified for a two-phase laminate with inelastic rate-(in)dependent phases and the quadratic convergence of the Newton solver has been confirmed for this benchmark. Its applicability for realistic problems has been demonstrated using a micrograph-based analysis of a sample of dual-phase steel with elasto-plastic phases.

Based on these results, we conclude that
\begin{enumerate}
  \item FFT-based solvers can be constructed using a similar variational basis as done for conventional Finite Element Methods,
  \item in consequence, constitutive routines developed for non-linear finite
  element formulations can be directly interfaced to FFT-based solvers, while
  keeping the computational efficiency of the FFT-based method,
  \item the only role of the (material-dependent) reference problem, central to
  the Lippmann-Schwinger approaches, is to ensure the convergence of the
  Richardson scheme used to solve the resulting system of linearized equations. This work proposes to use other linear solvers instead, such as the Conjugate Gradient method, that rely on the (material-independent) projection matrix.
\end{enumerate}

As the next step, we will extend the presented developments to a finite-strain setting, departing from the recent works by Eisenlohr~\etal~\cite{eisenlohr_spectral_2013} and Kabel~\etal~\cite{kabel_efficient_2014}.

\section*{Acknowledgement}

Chaowei Du (Eindhoven University of Technology) is gratefully acknowledged for providing the micrograph of Figure~\ref{fig:ex-dp_experiment} and Milan Jirásek~(Czech Technical University in Prague) for his helpful critical comments on the manuscript. Jaroslav Vond\v{r}ejc was partially supported by the Czech Science Foundation under project No.~13-22230S and Tom de Geus was supported by the Materials innovation institute M2i, The Netherlands, under project number M22.2.11424.



\appendix

\section{Operators}\label{app:operators}

For the non-zero frequency $\Tk \in \xZd
\backslash \{ \T{0} \}$, the Fourier transform of the fourth-order projection
operator $\FT{\TG}$, introduced in~\eqref{eq:projection}, is provided by,
e.g.~\cite[Section~6]{milton_variational_1988},
\begin{align}
\FT{G}_{ijlm}(\Tk)
= &
\frac{1}{2}
\frac{
\Fz{i} \delta_{jl} \Fz{m}
+
\Fz{i} \delta_{jm} \Fz{l}
+
\Fz{j} \delta_{il} \Fz{m}
+
\Fz{j} \delta_{im} \Fz{l}}{\norm{\T{\xi}(\Tk)}^2}
\nonumber \\
& -
\frac{
\Fz{i} \Fz{j} \Fz{l} \Fz{m}}{
\norm{\T{\xi}(\Tk)}^4}
\label{eq:FT_projection}
\end{align}
where the scaled frequencies
$\xi_i$ account for the size of the unit cell through $\Fz{i} = k_i / L_i$ and $\delta_{ij}$ stands for the Kronecker delta. For
$\Tk = \T{0}$, $\FT{G}_{ijlm}(\T{0}) = 0$ because of the zero-mean property.

The Fourier transform of the Green operator $\TGamma\aux$, from
Eq.~\eqref{eq:L-S}, associated with the reference stiffness $\T{C}\aux$ is more
involved, e.g. \cite[Section~5.2]{michel_effective_1999}. For $\Tk \not =
\T{0}$, we assemble the second-order acoustic tensor
\begin{align}
A_{il}(\Tk)
=
\sum_{j,m=1}^2
C\aux_{ijlm}
\Fz{j} \Fz{m}
\end{align}
to express the fourth-order Green operator in the form
\begin{align}\label{eq:Gamma_explicit}
\FT{\Gamma}\aux_{ijlm}(\Tk)
=
\frac{1}{4}
\Bigl( &
A^{-1}_{jm}(\Tk)
\Fz{i} \Fz{l}
+
A^{-1}_{jl}(\Tk)
\Fz{i} \Fz{m}
\nonumber \\
& +
A^{-1}_{im}(\Tk)
\Fz{j} \Fz{l}
+
A^{-1}_{il}(\Tk)
\Fz{j} \Fz{m}
\Bigr).
\end{align}
For $\Tk = \T{0}$, we set again
$\FT{\Gamma}\aux_{ijlm}(\T{0}) = 0$. A direct calculation then reveals that
the two operators coincide for $C\aux_{ijlm} = ( \delta_{il} \delta_{jm} +
\delta_{im} \delta_{jl} )/2$.

\section{Matrix notation}\label{app:matrices}

\begin{figure}[ht]
 \centering
 \def\svgwidth{.5\textwidth}
 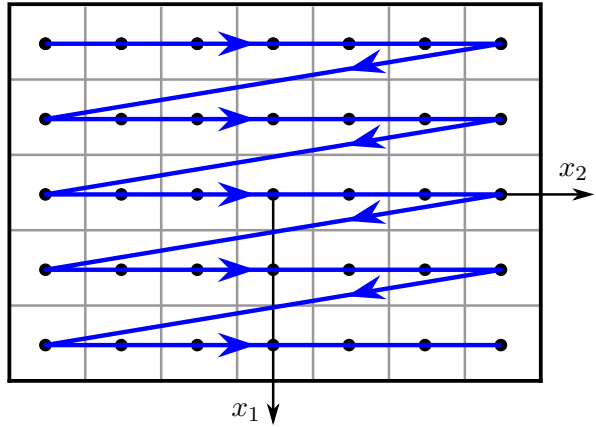
 \caption{Scheme of the vectorization operation.}
 \label{fig:finite_lattice_vector}
\end{figure}

On a regular grid $\xZNd$ with $\meas{\TN}$ nodes $\gpoint{\Tk}$, any periodic symmetric second-order trigonometric polynomial $\Tt$ and its Fourier
transform $\FT{\Tt}$ can be represented by the real- and complex-valued columns,
recall~\eqref{eq:matrix_representation_def},
\begin{align*}
\Mt
=
\left[
\begin{bmatrix}
\St_{11} \\
\St_{22} \\
\sqrt{2} \St_{12}
\end{bmatrix}
(\gpoint{\Tk})
\right]_{\Tk \in \xZNd}
\in \xRNdd,
&&
\FT{\Mt}
=
\left[
\begin{bmatrix}
\FT{\St}_{11} \\
\FT{\St}_{22} \\
\sqrt{2} \FT{\St}_{12}
\end{bmatrix}
(\Tk)
\right]_{\Tk \in \xZNd}
\in \xCNdd,
\end{align*}
where we have employed the Mandel representation,
e.g.~\cite[Section~2.3]{Milton:2002:TOC}. During this vectorization procedure,
data indexed by $\Tk \in \xZNd$ are gathered according to
\figref{fig:finite_lattice_vector}.

Under such nomenclature, the matrices from~\eqref{eq:DFT_matrices} implementing
the forward and the inverse Fourier transforms attain the form
\begin{align*}
\MF
=
\frac{1}{\meas{\TN}}
\begin{bmatrix}
\DFT{-\Tk}{\Tm}
\M{I}_{(3 \times 3)}
\end{bmatrix}_{\Tk, \Tm \in \xZNd}
\in
\xC^{3 \meas{\TN} \times 3 \meas{\TN}},
&&
\MF\inv
=
\begin{bmatrix}
\DFT{\Tk}{\Tm}
\M{I}_{(3 \times 3)}
\end{bmatrix}_{\Tk, \Tm \in \xZNd}
& \in
\xC^{3 \meas{\TN} \times 3 \meas{\TN}},
\end{align*}
where $\M{I}_{(3 \times 3)}$ is the $3\times 3$ unit matrix. The Fourier
transform of the projection matrix $\FT{\MG}$,
Eq.~\eqref{eq:discrete_projection}, is obtained as
\begin{align*}
\FT{\MG}
& =
\left[
\delta^{\Tk \Tm}
\begin{bmatrix}
\FT{G}_{1111} &
\FT{G}_{1122} &
\sqrt{2} \FT{G}_{1112}
\\
\FT{G}_{1122} &
\FT{G}_{2222} &
\sqrt{2} \FT{G}_{2212} \\
\sqrt{2} \FT{G}_{1112} &
\sqrt{2} \FT{G}_{2212} &
2 \FT{G}_{1212}
\end{bmatrix}
(\Tk)
\right]_{\Tk, \Tm \in \xZNd}
\in
\xR^{3 \meas{\TN} \times 3 \meas{\TN}},
\end{align*}
with $\delta^{\Tk \Tm}$ standing again for the Kronecker delta. Likewise, the
matrix form of the Green operator from~\eqref{eq:L-S_discrete} reads
\begin{align*}
\FT{\M{\Gamma}}\aux
& =
\left[
\delta^{\Tk \Tm}
\begin{bmatrix}
\FT{\Gamma}\aux_{1111} &
\FT{\Gamma}\aux_{1122} &
\sqrt{2} \FT{\Gamma}\aux_{1112}
\\
\FT{\Gamma}\aux_{1122} &
\FT{\Gamma}\aux_{2222} &
\sqrt{2} \FT{\Gamma}\aux_{2212}
\\
\sqrt{2} \FT{\Gamma}\aux_{1112} &
\sqrt{2} \FT{\Gamma}\aux_{2212} &
2 \FT{\Gamma}\aux_{1212}
\end{bmatrix}
(\Tk)
\right]_{\Tk, \Tm \in \xZNd}
\in
\xR^{3 \meas{\TN} \times 3 \meas{\TN}}.
\end{align*}

The conversion to the matrix format is completed by the treatment of the
constitutive laws. Specifically, the stresses
from~\eqref{eq:matrix_constitutive_law} need to be arranged in a column
\begin{align*}
\Mstress
=
\left[
\begin{bmatrix}
\stress_{11} \\
\stress_{22} \\
\sqrt{2} \stress_{12}
\end{bmatrix}
\left(\gpoint{\Tk}, \TE + \Tstrainfl(\gpoint{\Tk}) \right)
\right]_{\Tk \in \xZNd}
\in \xRNdd,
\end{align*}
whereas the tangent matrix~\eqref{eq:local_tangent} attains the form of a
block-diagonal $3 \meas{\TN} \times 3 \meas{\TN}$ matrix:
\begin{align*}
\M{C}
=
\left[
\delta^{\Tk \Tm}
\begin{bmatrix}
\partial\stress_{11} / \partial\strain_{11} &
\partial\stress_{11} / \partial\strain_{22} &
\sqrt{2} \partial\stress_{11} / \partial\strain_{12}
\\
\partial\stress_{22} / \partial\strain_{11} &
\partial\stress_{22} / \partial\strain_{22} &
\sqrt{2} \partial\stress_{22} / \partial\strain_{12}
\\
\sqrt{2} \partial\stress_{12} / \partial\strain_{11} &
\sqrt{2} \partial\stress_{12} / \partial\strain_{11} &
2 \partial\stress_{12} / \partial\strain_{12}
\end{bmatrix}
\left(\gpoint{\Tk}, \TE + \Tstrainfl(\gpoint{\Tk}) \right)
\right]_{\Tk, \Tm \in \xZNd}.
\end{align*}

Finally, the spaces of the nodal values of general, $\xTN$, and compatible, $\xEN$, trigonometric polynomials from \secref{sec:discretization} are provided by
\begin{align}\label{eq:discrete_spaces}
\xTN = \xRNdd, &&
\xEN = \MF\inv \FT{\MG} \, \MF
\left[ \xRNdd \right].
\end{align}

\end{document}